\documentclass[
reprint,
superscriptaddress,
amsmath,amssymb,
aps,
prl,
floatfix,
]{revtex4-2}

\usepackage{graphicx}
\usepackage{dcolumn}
\usepackage{bm}
\usepackage{extarrows}
\usepackage{xurl}
\usepackage[colorlinks = true,
            linkcolor = blue,
            urlcolor  = blue,
            citecolor = blue,
            anchorcolor = blue]{hyperref}
\usepackage{makecell}
\usepackage{bbding}

\newcommand{\parhead}[1]{\textit{#1}---}

\begin{document}

\title{Excitation density controlled regimes of collective light--matter dynamics}

\author{Wenxiang Ying}
\email{wying3@sas.upenn.edu}
\affiliation{Department of Chemistry, University of Pennsylvania, Philadelphia, Pennsylvania 19104, USA}

\author{Abraham Nitzan}
\email{anitzan@sas.upenn.edu}
\affiliation{Department of Chemistry, University of Pennsylvania, Philadelphia, Pennsylvania 19104, USA}
\affiliation{School of Chemistry, Tel Aviv University, Tel Aviv 69978, Israel}

\begin{abstract}
Theoretical descriptions of collective light--matter dynamics often rely on the mean-field (MF) or single-excitation (SE) approximations, yet the parameter regimes where they apply are rarely clearly delineated. 
Here we show that representative limiting regimes are characterized by two independent parameters: the number of molecules $N$ and the excitation number $N_{\rm exc}$.
In the Tavis--Cummings model, when $N\gg 1$ and the excitation density $N_{\rm exc} / N \to 0$, MF and SE descriptions agree and yield linear collective dynamics, showing harmonic Rabi oscillations. 
At finite excitation density ($N_{\rm exc} / N \sim \mathcal{O}(1)$), the large-\(N\) limit remains accurately described by MF dynamics but becomes nonlinear in $N_{\rm exc} / N$, manifested by a Duffing equation for the cavity amplitude with anharmonic Rabi frequency. 
We further show that cluster expansion systematically restores finite-$N$ correlations beyond MF. When local vibronic interactions are included, the same linear collective limit is reached by both approximations, with SE reaching it through polaron decoupling and MF through linearization.
This two-parameter regime map clarifies the limits in which different theoretical descriptions provide controlled descriptions of collective light--matter dynamics.
\end{abstract}

\maketitle

\begin{figure}[t]
  \includegraphics[width=\columnwidth]{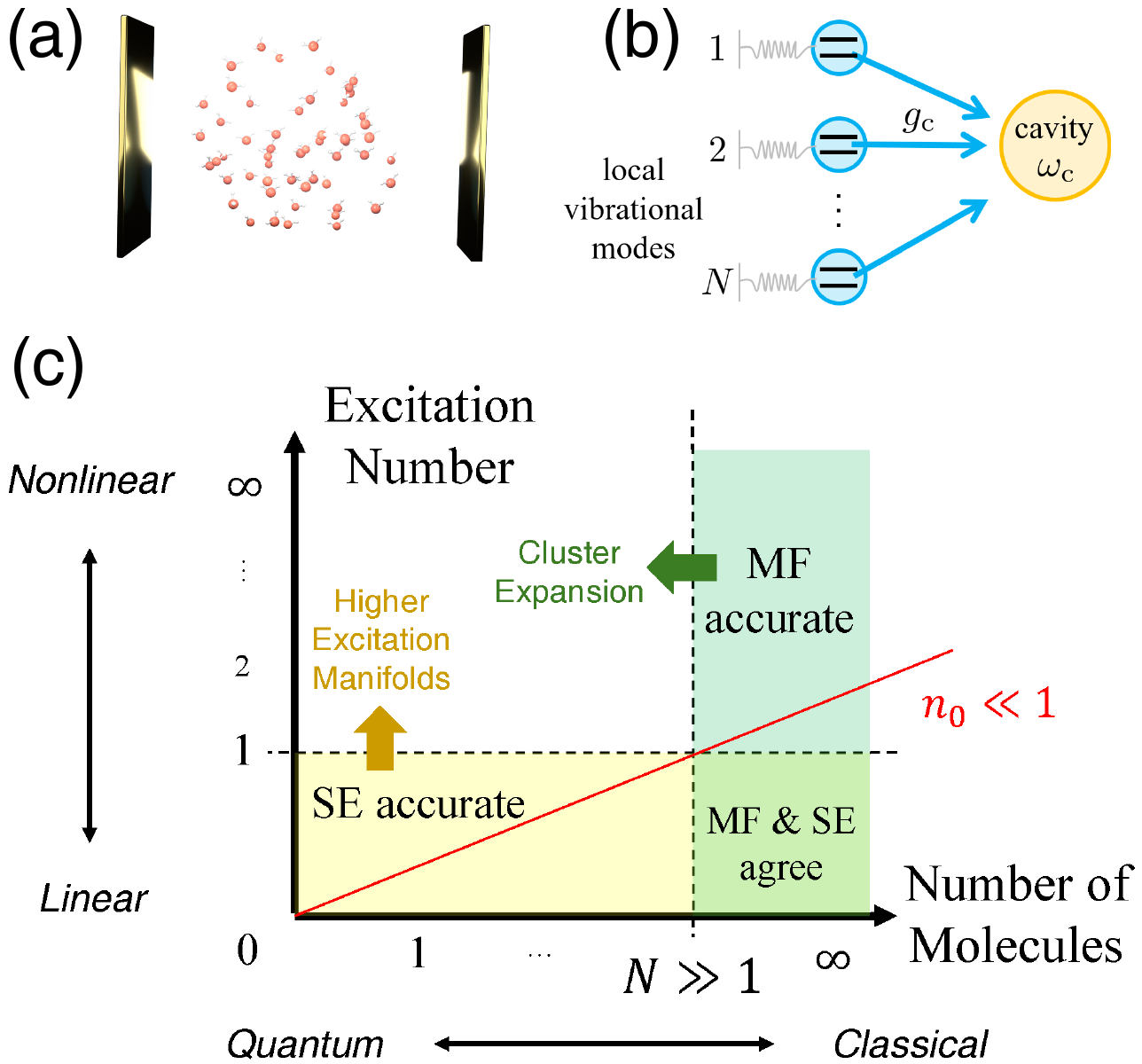}
  \caption{The HTC model and dynamical regime map.
  (a)~Schematic of a molecular ensemble coupled to a single cavity mode.
  (b)~Coupling topology: the cavity mode couples collectively to electronic transitions
  with single-molecule strength $g_\mathrm{c}$ (collective Rabi splitting
  $\Omega = 2g_\mathrm{c}\sqrt{N}$); each electronic excitation is locally coupled
  to its own vibration with strength $c_\nu$.
  (c)~Schematic two-parameter map in the $(N,\,N_\mathrm{exc})$ plane.
  The red curve marks constant excitation density $n_0 = N_\mathrm{exc}/N$; dashed
  lines indicate qualitative crossovers rather than sharp boundaries.
  Increasing $N$ drives the quantum-to-classical crossover, whereas increasing $n_0$
  drives the linear-to-nonlinear crossover.
  Representative limiting cases are summarized in Table~\ref{tab:regimes}.}
  \label{fig:schematic}
\end{figure}

\parhead{Introduction}
Molecular polaritons---hybrid light--matter states formed by strong coupling between
molecular ensembles and a confined electromagnetic mode (Fig.~\ref{fig:schematic}a)---are
routinely described by two theoretical frameworks that are frequently compared as
alternatives: the semiclassical mean-field (MF) approximation and the quantum single-excitation (SE) approximation.
MF assumes a single-configuration (product state) wavefunction for the cavity field and the molecular subsystems~\cite{Cui_JCP2022, Cui_JCP2023, Keeling_PRL2022, Hsieh_JCP2023, Hsieh_JPCL2023, Reichman_Nanoph2024,
Li_CP2025, Yuen-Zhou_PRL2025, Hsieh_JCTC2025}.
SE restricts the Hilbert space to the subspace of the ground plus single excitation states, enabling an exact
quantum treatment within this subspace~\cite{Herrera_Spano_PRL, herrera2018theory, Lai_JCP2024, Hu_JCP2025, Lai_JCP2026, Chng_2024, Ying_mn24, Chng_NL2025, Ying_NC2025}.
Both approximations are widely used as computationally efficient descriptions of polariton dynamics and can yield consistent results in selected applications. However, the regimes in which each approximation is controlled, and the conditions under which they agree, remain unclear.

Here we show that the relevant regimes can be characterized by two independent parameters. First, the number of molecules \(N\), which controls the suppression of quantum fluctuations in collective variables~\cite{Carollo_PRL2021,
Carollo_NJP2023} and hereby the approach to classical MF dynamics.
Second, the excitation number $N_{\rm exc}$ or the excitation density \(n_0=N_{\rm exc}/N\), which controls the validity of weak-excitation
linearization: $n_0\ll1$ keeps the dynamics linear while a finite $n_0$ produces nonlinearity.
Representative limiting regimes of the collective dynamics are summarized in Fig.~\ref{fig:schematic}c and Table~\ref{tab:regimes}.
When $N \gg 1$ and $n_0 \to 0$, MF and SE agree and yield linear collective dynamics, recovering harmonic Rabi oscillations for the Tavis--Cummings model;
with $N \gg 1$ but finite $n_0 \sim \mathcal{O}(1)$, the accuracy of MF description is preserved but becomes nonlinear and may be described by a Duffing equation for the cavity amplitude.
We further investigate less familiar situations: First, we use cluster expansion (CE)~\cite{Fricke_1996, Kira_2006, Koch_PRA2008, Ritsch_2022, Fowler-Wright_2024} to recover finite-$N$ correlations beyond MF approximation.
Secondly, we show that in the presence of local vibrations and vibronic interactions, modeled through the Holstein--Tavis--Cummings (HTC)
Hamiltonian~\cite{Li2021ARPC, Arkajit_Chemrev_2023, Ying_ARPC2026}, both MF and SE reach the same linear collective ($n_0 \to 0$) limit, with SE reaching it through polaron decoupling and MF through linearization. providing a regime map in terms of $(N, N_\mathrm{exc})$ for collective light--matter dynamics.

\parhead{Model and dynamical regimes}
The HTC Hamiltonian~\cite{Li2021ARPC, Arkajit_Chemrev_2023, Ying_ARPC2026}
describes $N$ identical two-level molecules, each with two electronic levels
$|g_n\rangle$ and $|e_n\rangle$ (for molecule $n$) associated with an energy gap $\hbar \omega_0$, and one local harmonic vibration with frequency $\nu$, collectively coupled to
a single cavity photon mode of frequency $\omega_\mathrm{c}$:
\begin{align}
  \hat{H}_\mathrm{HTC}
  &= \hbar\omega_\mathrm{c}\hat{a}^\dagger\hat{a}
   + \hbar\sum_{n=1}^{N}\Bigl[
     \omega_0\hat{\sigma}_n^+\hat{\sigma}_n^-
     + \nu\,\hat{b}_n^\dagger\hat{b}_n
     \notag\\
  &\quad + c_\nu\hat{\sigma}_n^+\hat{\sigma}_n^-\!\left(\hat{b}_n^\dagger + \hat{b}_n\right)
   + g_\mathrm{c}\!\left(\hat{a}^\dagger\hat{\sigma}_n^- + \hat{a}\hat{\sigma}_n^+\right)
   \Bigr],
  \label{eq:HTC}
\end{align}
where $\hat{\sigma}_n^-$ ($\hat{\sigma}_n^+$) are the annihilation (creation) operators for the excitation of the $n_\mathrm{th}$ molecule, $\hat{b}_{n}$ ($\hat{b}^\dagger_{n}$) are the annihilation (creation) operators for the vibrational mode on the $n_\mathrm{th}$ molecule, and $c_\nu$ represents the molecular vibronic coupling strength, sometimes represented by the Huang-Rhys factor $S = (c_{\nu} / \nu)^2$.
Furthermore, $\hat{a}$ ($\hat{a}^\dagger$) are the annihilation (creation) operators for the cavity photon mode. The single-molecular light-matter coupling strength is $g_\mathrm{c}$. Fig.~\ref{fig:schematic}b shows schematics of the HTC Hamiltonian.
Unless otherwise specified, we focus on the resonant light-matter coupling, with $\omega_\mathrm{c}=\omega_0$.

Two parameters characterize the qualitative regime map of the HTC model.
The first is the number of molecules $N$ --- quantum fluctuations of collective variables like $\hat{S}^\pm = N^{-1}\sum_k\hat{\sigma}_k^\pm$ are suppressed as $1/\sqrt{N}$ and vanish as $N\to\infty$, rendering MF exactness for collective observables~\cite{Carollo_PRL2021, Carollo_NJP2023}.
The second parameter is the excitation number $N_{\rm exc}$, where the density $n_0 = N_{\rm exc}/N$ controls the weak-excitation linearization in the large-$N$ limit 
(see the Duffing equation~\eqref{eq:Duffing} below).
Table~\ref{tab:regimes} summarizes representative limiting cases organized by $N$ and
$N_{\rm exc}$, together with model examples that belong to the corresponding limits.
Fig.~\ref{fig:schematic}c shows the same information as a schematic regime map; the
dashed lines indicate qualitative crossovers rather than sharp thresholds.

\begin{table*}[t]
\centering
\caption{Representative limiting regimes of collective light--matter dynamics.
Checkmarks indicate that the corresponding property holds in the given limit.
JC: Jaynes--Cummings ($N=1$); TC: Tavis--Cummings ($N\gg1$).}
\label{tab:regimes}
\small
\renewcommand{\arraystretch}{1.35}
\begin{tabular*}{\textwidth}{@{\extracolsep{\fill}}ccccccccc}
\hline\hline
\multicolumn{2}{c}{Parameters} & excitation density & \multicolumn{3}{c}{Dynamical regime} & \multicolumn{2}{c}{Reliability} & Representative \\
\cline{1-2}\cline{4-6}\cline{7-8}
$N$ & $N_\mathrm{exc}$ & $n_0 = N_\mathrm{exc}/N$ & Collectivity & Linearity & Classicality & MF & SE & example \\
\hline
Small & Small &
  $\mathcal{O}(1)$ &
  $\times$ & \checkmark & $\times$ &
  $\times$ & \checkmark &
  Singly excited JC \\
Large & Small &
  $\mathcal{O}(1/N)$ &
  \checkmark & \checkmark & \checkmark &
  \checkmark & \checkmark &
  Weakly excited TC \\
Large & Large &
  $\mathcal{O}(1)$ &
  \checkmark & $\times$ & \checkmark &
  \checkmark & $\times$ &
  Multiply excited TC \\
Small & Large &
  $\gtrsim\mathcal{O}(1)$ &
  $\times$ & $\times$ & $\times$ &
  $\times$ & $\times$ &
  Multiply excited JC \\
\hline\hline
\end{tabular*}
\end{table*}

Note that the entries in Table~\ref{tab:regimes} use three related but distinct notions.
\textit{Collectivity} refers to $N \gg 1$ ensemble-bright
degrees of freedom, such as $\hat{B} = N^{-1/2}\sum_n\hat{\sigma}_n^-$ or
$\hat{S}^\pm = N^{-1}\sum_n\hat{\sigma}_n^\pm$ in the $N \gg 1$ limit and does not by itself imply classicality~\cite{Schwengelbeck_2026}.
\textit{Linearity} refers to the consequence of linearizing the equations of motion about the weak-excitation limit ($n_0\ll1$). In the Tavis--Cummings model considered
below, this linearized dynamics yields harmonic Rabi oscillations, but linearity and
harmonicity are not synonymous in general.
\textit{Classicality} refers to the suppression of quantum fluctuations as $N\to\infty$, enabling a reliable MF description (in terms of single-operator expectation values) for collective observables, such as $\alpha(t)$,
$\sigma(t)$, and $w(t)$ in Eq.~\ref{eq:MBE}. 
In this limit, connected higher-order correlations are suppressed, so the quantum Bogoliubov-Born-Green-Kirkwood-Yvon (BBGKY) hierarchy~\cite{Cox_PRA2018, Limmer_2024} admits a MF closure at the first-order.

In the following sections we demonstrate this regime-structure analytically and
numerically, starting from the bare Tavis--Cummings (TC) model~\cite{tavis1968exact, Tavis_1969} ($c_\nu = 0$) and then
extending to the full HTC model.

\parhead{Nonlinear mean-field dynamics and the Duffing equation}
We first analyze the TC model ($c_\nu = 0$) to expose the role of $n_0$ without vibronic interactions.
The dark modes decouple exactly from the cavity, and 
the TC model reduces to an effective two-mode problem between the cavity and the
collective bright mode $\hat{\sigma}_\mathrm{B}^\pm = N^{-1/2}\sum_n\hat{\sigma}_n^\pm$.
In the SE limit, this gives two quantum polariton states with collective Rabi
splitting $\Omega = 2g_\mathrm{c}\sqrt{N}$~\cite{tavis1968exact, Tavis_1969}.

The MF approximation yields the semiclassical Maxwell--Bloch equations~\cite{Rose2017, Angerer2017, Keeling_PRL2022} for the cavity amplitude
$\alpha = \langle\hat{a}\rangle$, the collective electronic coherence
$\sigma = N^{-1}\sum_n\langle\hat{\sigma}_n^-\rangle$, and the mean inversion
$w = N^{-1}\sum_n\langle\hat{\sigma}_n^z\rangle \in [-1,1]$~\footnote{Although
Eqs.~\ref{eq:MBE} have the algebraic structure of the externally driven Bloch
equations, a crucial difference is that $\alpha(t)$ is itself a dynamical variable
whose back-action on the matter equations produces amplitude-dependent (anharmonic)
Rabi oscillations rather than the strictly harmonic precession of externally driven
spins.}:
\begin{subequations}
\label{eq:MBE}
\begin{align}
  \dot{\alpha} &= -i\omega_\mathrm{c}\alpha - ig_\mathrm{c} N \sigma,
  \label{eq:MBE-1} \\
  \dot{\sigma} &= -i\omega_0\sigma + ig_\mathrm{c} \alpha\, w,
  \label{eq:MBE-2} \\
  \dot{w}      &= 2ig_\mathrm{c}\!\left(\alpha^* \sigma - \alpha \sigma^*\right).
  \label{eq:MBE-3}
\end{align}
\end{subequations}
See Supporting Information, Sec. I-A for derivations.
In the weak-excitation limit ($n_0 \to 0$, $w \approx -1$), Eqs.~\ref{eq:MBE}
linearize to $\dot{\alpha} = -i\omega_\mathrm{c}\alpha - ig_{\mathrm{c}}N\sigma$
and $\dot{\sigma} = -i\omega_0\sigma - ig_{\mathrm{c}}\alpha$.
The normal-mode frequencies
$\omega_\pm = \tfrac{1}{2}(\omega_\mathrm{c}+\omega_0) \pm \sqrt{g_{\mathrm{c}}^2 N
+ \tfrac{1}{4}(\omega_\mathrm{c}-\omega_0)^2}$
are identical to the quantum polariton eigenenergies~\cite{Li2021ARPC, Arkajit_Chemrev_2023, Ying_ARPC2026}, confirming that large-$N$ MF
recovers SE in the linear collective regime. 

Beyond the weak-field regime, the nonlinear term $ig_\mathrm{c}\alpha w$ drives $w$
away from $-1$.
For simplicity, we focus on the initial condition $\alpha(0) = \alpha_0 \in \mathbb{R}$, $\sigma(0) = 0$,
$w(0) = -1$. The excitation number is $N_{\rm exc} = |\alpha(0)|^2$, and conservation of total excitation
$\frac{d}{dt}[|\alpha|^2 + \tfrac{N}{2}(1+w)] = 0$ yields
$w = -1 + 2(n_0 - |\alpha|^2/N)$ with $n_0 = |\alpha_0|^2/N$ the excitation
density.
In the resonant rotating frame ($\omega_\mathrm{c} = \omega_0$), differentiating
Eq.~\ref{eq:MBE-1} and substituting yields the Duffing equation~\cite{Hsieh_JCP2023}
\begin{align}
  \ddot{\alpha} + g_\mathrm{c}^2 N(1 - 2n_0)\,\alpha + 2g_\mathrm{c}^2\,\alpha^3 = 0,
  \label{eq:Duffing}
\end{align}
where the cubic term reflects Bloch-sphere curvature and $|\alpha|^2\alpha \to \alpha^3$
because the rotating-frame dynamics preserves $\alpha(t) \in \mathbb{R}$ for the chosen
initial condition.
Derivation details are in Supporting Information, Sec.~I.

Eq.~\ref{eq:Duffing} can be solved exactly by separation of variables and yields a solution in terms of Jacobi elliptic functions. Details are presented in Supporting Information, Sec.~II. 
A harmonic trial solution $\alpha(t) = \alpha_0\cos(\tilde\omega t)$, retaining
only the fundamental frequency component, gives the effective Rabi frequency that depends on excitation density $n_0$,
\begin{align}
  \Omega^\mathrm{eff} \approx 2g_\mathrm{c}\sqrt{N}\!\left(1 - \frac{n_0}{4}\right),
  \label{eq:Rabi_eff}
\end{align}
to leading order in $n_0 \to 0$.
As $n_0 \to 0$, $\Omega^\mathrm{eff}$ recovers the collective Rabi splitting $\Omega$~\footnote{Note that large $N$ with $n_0 = \mathcal{O}(1)$ reflects the saturation of excitation in MF rather than breakdown. Eq.~\ref{eq:Duffing} is the exact large-$N$ dynamics, arising because MF becomes exact for collective observables in the thermodynamic limit~\cite{Carollo_PRL2021, Carollo_NJP2023}. The harmonic SE-like Rabi dynamics requires the additional limit $n_0 \to 0$.}.

\begin{figure}[t]
  \includegraphics[width=\columnwidth]{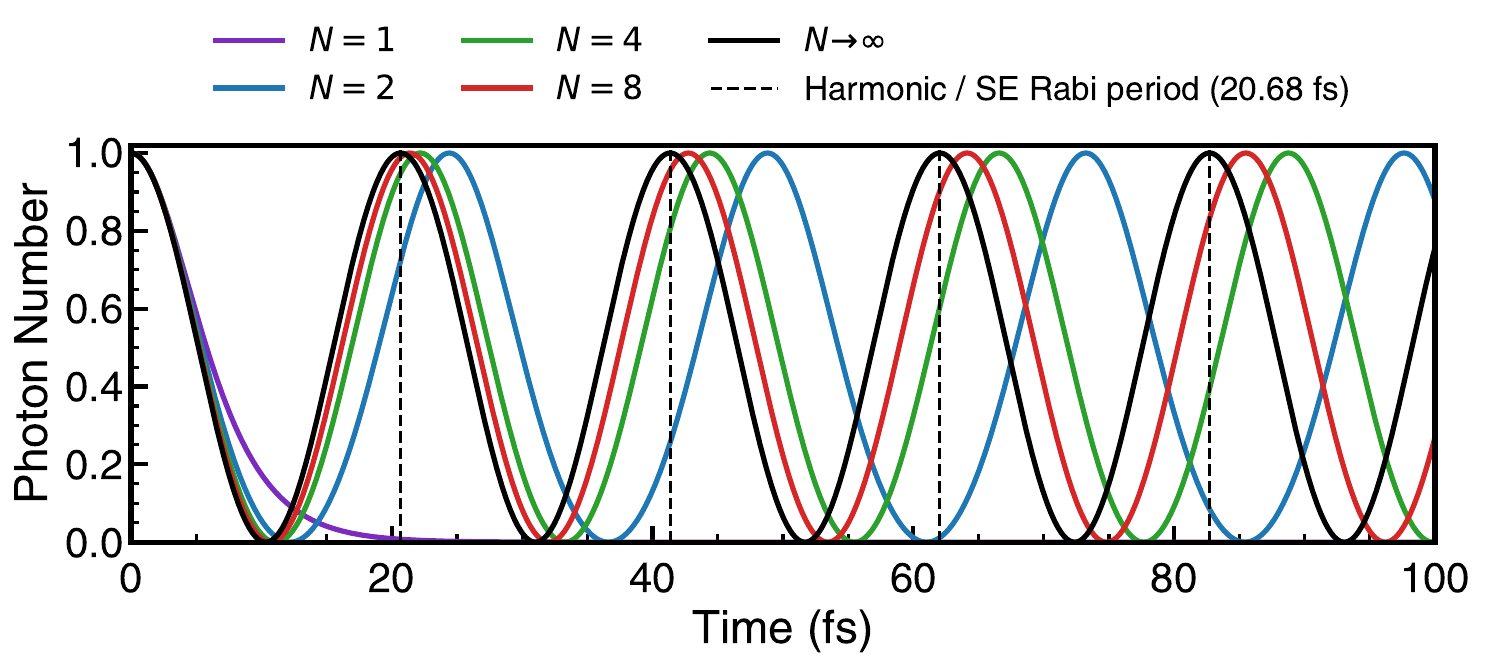}
  \caption{Photon number dynamics $|\alpha(t)|^2$ of the TC model ($c_\nu = 0$) calculated with MF for
  $N = 1, 2, 4, 8, 10^4$ with fixed $\alpha_0 = 1$, so $N_{\rm exc} \equiv |\alpha_0|^2 = 1$ and $n_0 = 1/N$ decreases
  with increasing $N$.
  The large-$N$ (small-$n_0$) limit converges to the harmonic SE Rabi oscillation
  (dashed line), demonstrating the linear collective regime ($N\gg1$, $n_0\ll1$).
  Anharmonic Duffing dynamics is evident at small $N$ (large $n_0$).}
  \label{fig:rabi_convergence}
\end{figure}

Fig.~\ref{fig:rabi_convergence} displays this picture --- it shows the MF photon number $|\alpha(t)|^2$, obtained by solving Eqs.~\ref{eq:MBE} for $N = 1, 2, 4, 8, 10^4$
with $\omega_0 = \omega_\mathrm{c} = 2.0$~eV and $\sqrt{N}g_\mathrm{c} = 0.10$~eV ($\Omega = 0.20$~eV, Rabi period
$T = 20.68$~fs) and fixed $\alpha_0 = 1$, so that $n_0 = 1/N$ decreases
as $N$ increases.
The dynamics converges monotonically to the harmonic SE Rabi oscillation as
$n_0 \to 0$ (large $N$), demonstrating the linear collective regime ($N\gg1$, $n_0 \to 0$).
At small $N$ ($n_0 \sim \mathcal{O}(1)$), the anharmonicity is clear, showing oscillation periods that deviate from the black dashed lines.
The convergence is controlled by $n_0$, not $N$ alone: if $|\alpha_0|^2$ were
scaled proportionally to $N$, the anharmonicity would persist for any $N$.

\parhead{Effect of higher order correlations}
The MF product-state closure is controlled for collective observables at large $N$,
but finite-$N$ fluctuations and quantum correlations are not captured.
Cluster expansion (CE)~\cite{Fricke_1996, Kira_2006, Koch_PRA2008, Ritsch_2022, Fowler-Wright_2024} provides a systematic framework for restoring finite-$N$ quantum correlations beyond MF, in which higher-order intermolecular correlations scale as $\mathcal{O}(1/N)$.
Details of the CE hierarchy, identity constraints, closure conditions, and numerical
implementations are given in Supporting Information, Sec.~III.
We note that the collective dynamics using truncated equations (CUT-E) approach~\cite{Joel_PNAS_23} developed by Yuen-Zhou, {\it et al.} also leverages the permutational symmetry of matter states through an effective $1/N$ expansion~\cite{Joel_JCP2025}. 

\begin{figure}[t]
  \includegraphics[width=\columnwidth]{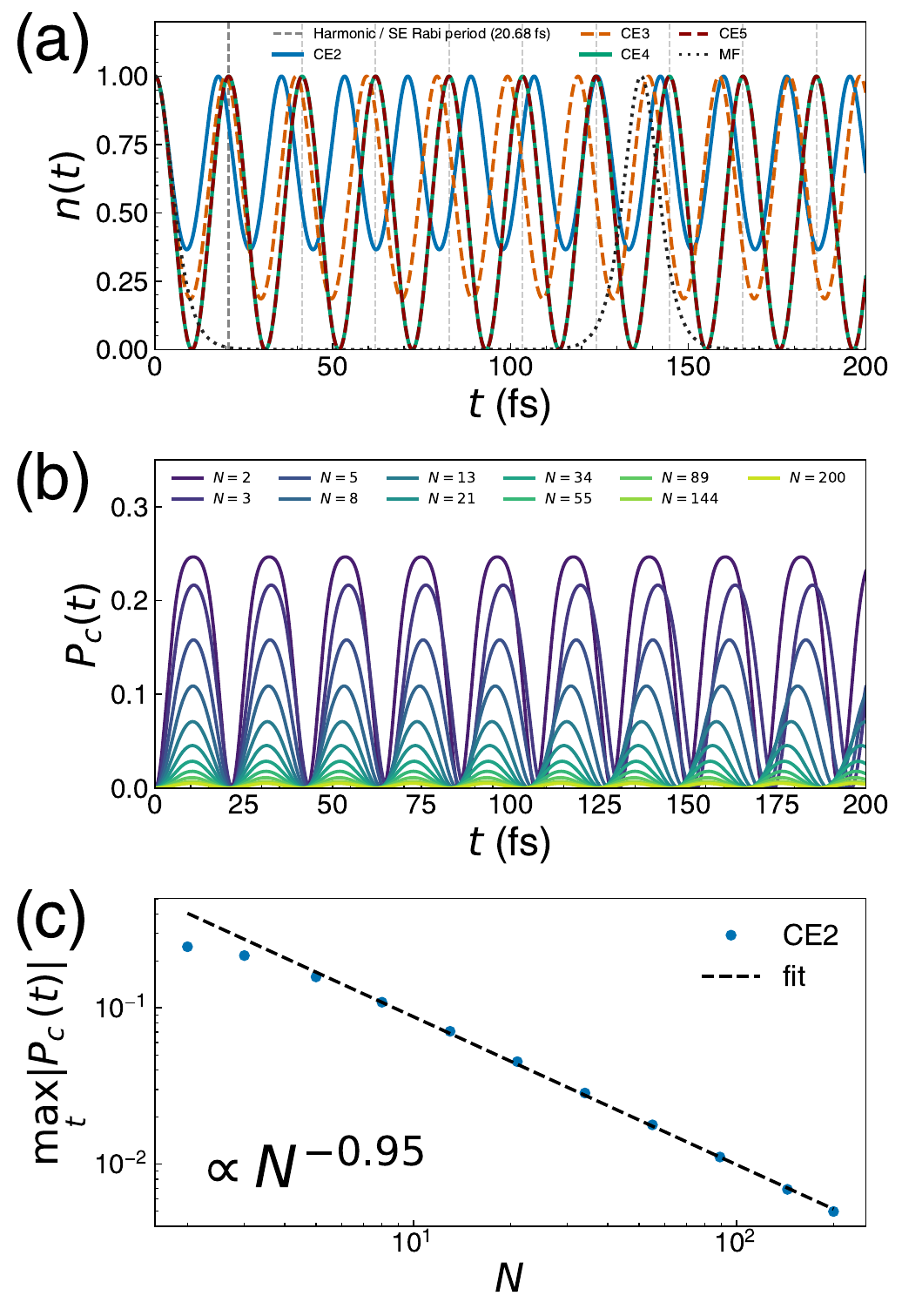}
  \caption{Cluster-expansion dynamics and finite-$N$ correlations.
  (a)~Photon number dynamics $n(t)$ from identity-constrained CE hierarchy (colored curves) for $N=1$,
  $N_\mathrm{exc}=1$, initialized from
  $|1\rangle\otimes|g\rangle$. The MF curve (black dots) uses the coherent-field counterpart with the same initial mean photon number, $|\alpha_0|^2=1$. 
  Vertical dashed lines mark the harmonic/SE
  Rabi period, $T=20.68$~fs.
  (b)~Two-molecule correlation
  $P_c=\langle\hat\sigma_i^+\hat\sigma_j^-\rangle
  -|\langle\hat\sigma_i^-\rangle|^2$ $(i\neq j)$ obtained using many-molecule CE2 for various
  $N$ under the same one-photon initial
  condition and fixed collective Rabi splitting $\Omega$ (so that $g_\mathrm{c} \sim 1 / \sqrt{N}$).
  (c)~Peak values of $|P_c(t)|$ versus $N$ with a black dashed line from power-law
  fit.}
  \label{fig:ce_convergence}
\end{figure}

Fig.~\ref{fig:ce_convergence}a presents the photon number dynamics $n(t) = \langle \hat{a}^\dagger \hat{a} \rangle (t)$ from identity-constrained CE dynamics for the JC model initialized in the fixed-excitation state $|1\rangle\otimes|g\rangle$; the MF comparison uses a coherent field with the same initial mean photon number ($\alpha_0 = 1$, the same as Fig.~\ref{fig:rabi_convergence}), with $n(t) = |\alpha(t)|^2$. 
Here CE\(k\) denotes a cumulant hierarchy truncated after
\(k\)-body connected correlations.  Because the exact dynamics is restricted to
the two-state manifold $\{|1,g\rangle,|0,e\rangle\}$, the higher-order CE
implementation must also enforce the corresponding operator identities.  With
these constraints, CE2--CE5 provide a controlled finite-$N$ correction to the
MF product-state closure, and the dashed vertical lines mark the harmonic/SE
Rabi period that characterizes the exact single-excitation exchange.  The
necessity of imposing the constraints is illustrated in
Supporting Information, Sec.~III-B and Fig.~S1.

Figs.~\ref{fig:ce_convergence}b-c turns to the many-molecule cases.
The many-molecule CE2 hierarchy keeps both field--molecule and intermolecular correlations.
Panel (b) shows the two-molecule correlation
$P_c= \langle\hat\sigma_i^+\hat\sigma_j^-\rangle -|\langle\hat\sigma_i^-\rangle|^2$ ($i \neq j$) for a fixed
collective Rabi splitting $\Omega$ and the same one-photon initial
condition. Owing to permutation symmetry, $P_c$ is identical for all
distinct molecule pairs $(i,j)$. Details are provided in Supporting
Information, Sec.~III-C.
One observes that the amplitude is largest for small $N$ and is
progressively suppressed as $N$ increases, reflecting the dilution of one
quantum of excitation over more molecules.  Panel (c) summarizes this trend by
plotting the peak connected correlation versus $N$; the dashed power-law fit
shows the systematic decay of the finite-$N$ two-body correction, consistent
with recovery of the MF product-state limit for collective observables.

Having demonstrated several dynamical regimes of the TC model, we now restore the vibrational degrees of freedom ($c_\nu \neq 0$) and ask whether this regime
structure survives with local vibronic coupling.

\parhead{Convergence of bright vibrational mode dynamics}
We next include vibrations and compare SE and MF in the regime where both should
describe the same collective bright-sector observable. 

The Holstein coupling in
Eq.~\ref{eq:HTC} is strictly site-local, $\hat{H}_\mathrm{e\text{-}ph} = \hbar c_\nu \sum_n \hat{\sigma}_n^+\hat{\sigma}_n^- \!(\hat{b}_n^\dagger + \hat{b}_n )$. Introducing collective bright exciton $\hat{B} = N^{-1/2}\sum_n\hat{\sigma}_n^-$ and bright vibration operators
$\hat{b}_B = N^{-1/2}\sum_n\hat{b}_n$, and projecting $\hat{H}_\mathrm{e\text{-}ph}$ onto this representation yields
\begin{equation}
    \hat{H}_\mathrm{e\text{-}ph} = \frac{\hbar c_\nu}{\sqrt{N}}\,\hat{B}^\dagger\hat{B}
    \bigl(\hat{b}_B+\hat{b}_B^\dagger\bigr) + \text{dark-mode terms},
\label{eq:bright_Holstein}
\end{equation}
see Supporting Information, Sec.~IV-A.
Permutation-symmetric delocalization over $N$ molecules suppresses the bright-sector
vibronic coupling from $c_\nu$ to $c_\nu/\sqrt{N}$ --- the polaron decoupling effect~\cite{Herrera_Spano_PRL, herrera2018theory}.
We use a minimal two-internal-state truncation~\cite{Cui_JCP2023} to reveal the vibronic dynamics under the SE description:
each molecule has two vibrational states $|a\rangle$ and $|b\rangle$ on top of the two electronic levels $|g\rangle$ and $|e\rangle$. 
Let the all-ground state be
$|G_0\rangle=\prod_{j=1}^{N}|g_j,a_j\rangle$. In SE, the relevant bright basis contains two excitonic states
\begin{subequations}
\label{eq:bright_two_state_basis}
\begin{align}
  |B_0\rangle
  &= \frac{1}{\sqrt{N}}\sum_{k=1}^{N}
     |e_k,a_k\rangle\!\prod_{j\neq k}|g_j,a_j\rangle,
  \label{eq:B0_def}\\
  |B_1\rangle
  &= \frac{1}{\sqrt{N}}\sum_{k=1}^{N}
     |e_k,b_k\rangle\!\prod_{j\neq k}|g_j,a_j\rangle,
  \label{eq:B1_def}
\end{align}
\end{subequations}
representing the collective electronic excitation with different vibrational quantum
on the excited molecule, and two one-photon states
$|C_0\rangle=|1_\mathrm{ph}\rangle\otimes|G_0\rangle$ and
$|C_1\rangle=|1_\mathrm{ph}\rangle\otimes
N^{-1/2}\sum_k |g_k,b_k\rangle\allowbreak\prod_{j\neq k}|g_j,a_j\rangle$, corresponding to a cavity photon with different vibrational quantum, respectively. In this basis, the two light-matter channels
\begin{equation}
    \langle C_0|\hat{H}_\mathrm{HTC}|B_0\rangle=\hbar g_\mathrm{c}\sqrt{N},\,\,\,\,\,\, \langle C_1|\hat{H}_\mathrm{HTC}|B_1\rangle=\hbar g_\mathrm{c},
\end{equation}
exposes asymmetry: $|C_0\rangle$ couples collectively to $|B_0\rangle$, whereas coupling between $|C_1\rangle$ and $|B_1\rangle$ is not collectively enhanced and $|C_1\rangle$ state decouples on the Rabi oscillation timescale as $N$ grows. 
The full $4\times4$ Hamiltonian matrix in the $\{|B_0\rangle,|B_1\rangle,|C_0\rangle,|C_1\rangle\}$ basis is provided in Supporting Information, Sec.~IV-B. As a result, the SE dynamics under the large-$N$ limit is captured by the effective three-state subspace $\{|B_0\rangle,|B_1\rangle,|C_0\rangle\}$.

The corresponding MF description uses a permutation-symmetric molecular state
$|\psi\rangle=c_{ga}|g,a\rangle+c_{ea}|e,a\rangle+c_{gb}|g,b\rangle
+c_{eb}|e,b\rangle$ coupled to the cavity amplitude $\alpha$.
The behavior of the MF equations in the weak-excitation limit, $n_0 \to 0$, is presented in Supporting Information, Sec.~IV-C.
In this limit, $c_{ga}\simeq 1$ and $c_{gb} \to 0$, leading in linear order to a closed set of equations for $(\alpha,c_{ea},c_{eb})$. After rescaling, $\tilde{\alpha}=\alpha/\sqrt{N}$, these equations are identical to the large-$N$
SE three-state equations with
$(\tilde{\alpha},c_{ea},c_{eb})\leftrightarrow(|C_0\rangle,|B_0\rangle,|B_1\rangle)$, showing that the agreement between MF and SE in the limit \(N \gg 1\) and \(n_0 \to 0\) extends to collective vibronic dynamics.

\begin{figure}[t]
  \includegraphics[width=\columnwidth]{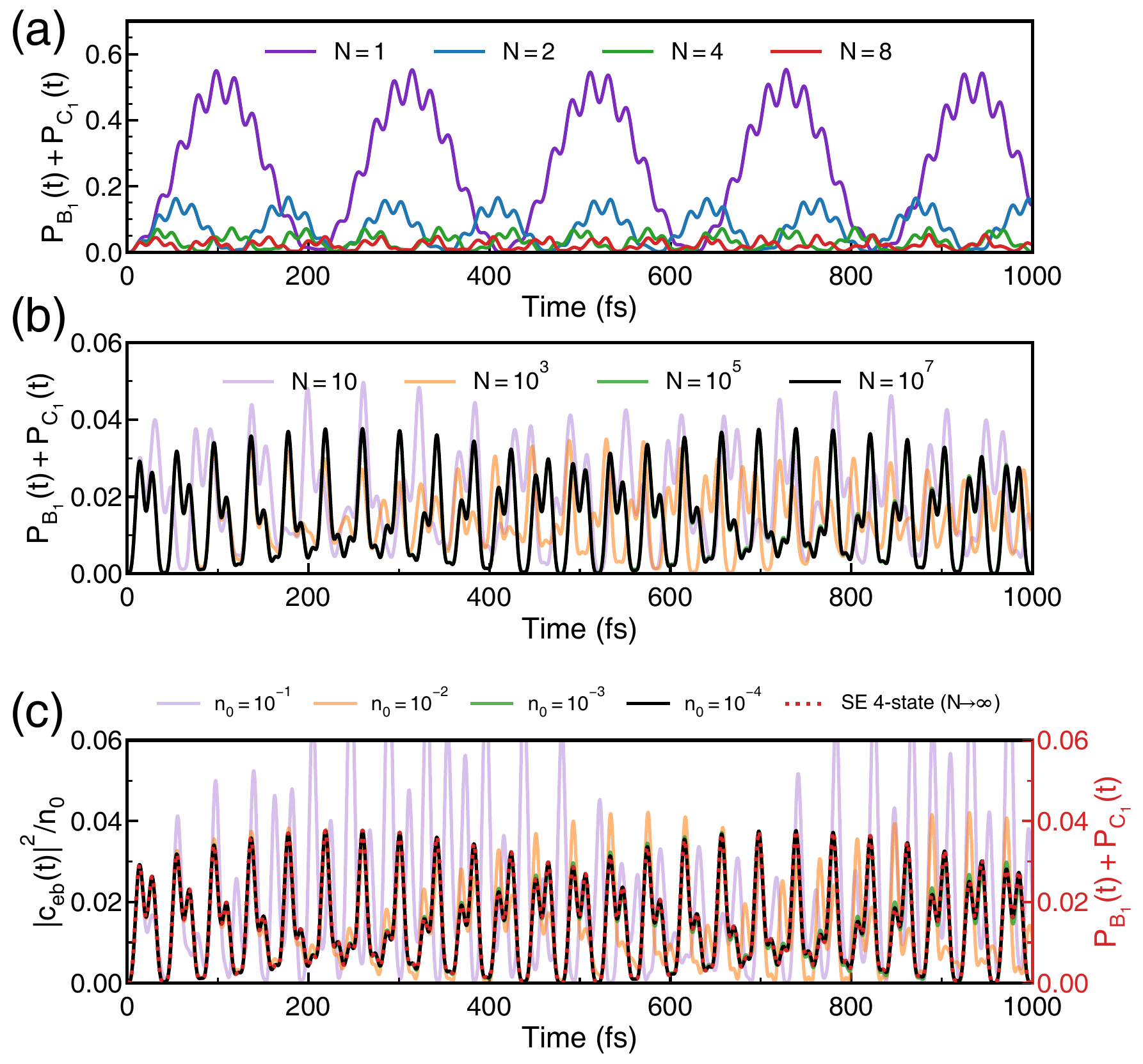}
  \caption{Convergence of the bright vibrational mode dynamics. Here we take $\nu = \Omega = 0.2$ eV and $c_\nu = 20$ meV.
  (a)~Vibrationally excited population $P_{B_1}(t)+P_{C_1}(t)$ for
  $N=1,2,4,8$ calculated in the SE approximation.
  (b)~Same as (a), for $N=10,10^3,10^5,10^7$, convergence reached as
  $N = 10^7$ (black curve).
  (c)~MF $b$-channel population $|c_{eb}(t)|^2/n_0$ for decreasing $n_0$ (colored
  curves) converges to the SE large-$N$ benchmark (red dashed curve), demonstrating
  MF--SE agreement in the linear collective regime. }
  \label{fig:se_mf_limit}
\end{figure}

Fig.~\ref{fig:se_mf_limit} illustrates this convergence, with $\nu = 0.2$~eV and $c_\nu = 20$~meV. Other parameters are identical to Fig.~\ref{fig:rabi_convergence} except that $N$ and $N_\mathrm{exc}$ are varied as convergence control parameters. 
Panels (a)--(b) show the population dynamics of the $b$-channel vibrational bright states,
$P_{B_1}(t)+P_{C_1}(t)$ with the SE and two-internal-state description ($N_\mathrm{exc} = 1$), first for few molecules ($N = 1,2,4,8$) and then across the large-$N$; increasing $N$ suppresses the oscillating amplitude and gradually converges to the black curve in panel (b) (with $N = 10^7$). 
Panel (c) shows that as $n_0 \to 0$, the MF $b$-channel population dynamics
$|c_{eb}(t)|^2/n_0$ gradually converges and approaches the large-$N$ SE benchmark (red dotted curve, identical to the black curve in panel (b)), confirming
agreement of the bright vibrational mode dynamics in the MF-SE overlap regime ($N \gg 1$, $n_0 \to 0$).
It is interesting to note that SE and MF converges agreement in the $N \to \infty$, $n_0 \to 0$ limit through different mechanisms: SE through polaron decoupling and MF through linearization of the equations of motion.

\parhead{Discussion}
The central result of this work is a two-parameter organization of collective
light--matter dynamics by molecule number $N$ and excitation number
$N_\mathrm{exc}$, or equivalently excitation density $n_0=N_\mathrm{exc}/N$.
Increasing $N$ suppresses collective quantum fluctuations and supports a
mean-field description of collective observables, whereas decreasing $n_0$
controls weak-excitation linearization.  Thus the linear collective regime
requires the combined limit $N\gg1$ and $n_0\to0$.  The regimes in
Table~\ref{tab:regimes} and Fig.~\ref{fig:schematic}c should be viewed as
representative limiting regimes connected by observable- and timescale-dependent
crossovers, rather than as sharp boundaries.

One caveat is that this large-\(N\) MF exactness is not generic to all many-body systems. It relies on the collective interaction topology of TC/HTC/Dicke-type models, where the coupling is mediated by global variables and scaled with system size so that the thermodynamic limit remains well behaved, analogous to Kac scaling in long-range interacting systems~\cite{Kac1963, Dauxois2002, Mori2010} and in the Lipkin--Meshkov--Glick model~\cite{Lipkin1965, Ribeiro2008}. 
By contrast, adding genuinely local interaction terms, such as short-range Hubbard-type couplings~\cite{Mandal_PRB2025}, can retain local correlations that are not suppressed by increasing $N$ and therefore need not admit an exact MF limit.

This distinction clarifies the domains of MF and SE descriptions.  Large
$N$ alone does not imply harmonic or linear dynamics: at finite excitation
density, the large-$N$ TC model remains accurately described by MF dynamics but becomes nonlinear in $n_0$, manifested by the Duffing equation~\eqref{eq:Duffing}.  The harmonic SE-like Rabi oscillation is
recovered only after the additional weak-excitation limit $n_0\to0$ is taken.
Conversely, at finite $N$, product-state MF misses quantum correlations that can
be restored systematically by cluster expansion. CE therefore provides a
controlled route from linear / nonlinear MF toward finite-$N$ quantum dynamics.

Local vibrational coupling does not destroy this overlap regime.  In the HTC
model, MF and SE still agree for collective bright-sector vibronic dynamics in
the joint limit $N\gg1$ and $n_0\to0$.  Importantly, this agreement arises
through different mechanisms: polaron decoupling in SE and weak-excitation
linearization in MF.  
Away from this overlap regime, the two descriptions should not be expected to remain equivalent: SE resolves finite-excitation quantum dynamics, whereas MF describes the thermodynamic response of collective observables and becomes nonlinear at finite excitation density.
The resulting $\sqrt{N}$ scaling of the resonant vibronic timescale~\cite{Ying_2026} provides a possible signature of polaron decoupling in ultrafast
pump--probe or multidimensional vibrational spectroscopy~\cite{Xiong_PNAS2018, xiong_Science2020, Xiong_Science2022, Son_CPR2024}.

More broadly, the $N$--$N_\mathrm{exc}$ regime map provides a diagnostic
framework for choosing and benchmarking controlled descriptions of collective
light--matter dynamics.  It can help revisit Dicke phase transitions and
superradiance by separating thermodynamic-limit MF behavior from finite-$N$
fluctuations; analyze driven-dissipative polariton dynamics where excitation
density and loss determine whether linearization is sufficient; organize
nonlinear spectroscopy and pump--probe experiments where finite excitation
density can move the system outside the SE regime; and benchmark CE,
truncated-Wigner, matrix-product-state, and multi-excitation quantum approaches
where neither MF nor SE alone is controlled.
Since classical electromagnetic simulations such as finite-difference time-domain (FDTD) effectively treat the optical field at the MF level~\cite{Sukharev_JCTC2025, Bustamante_JCTC2025, Sidler_Nanoph2025, bustamante_2026}, the present regime map also helps clarify their expected domain of validity.
Extensions to disordered ensembles~\cite{Xiong_2025} and multimode cavities~\cite{Ying2023NanoP} are natural next steps.\\

\parhead{Supporting Information}
The Supporting Information includes derivations of the Maxwell--Bloch equations,
the exact Duffing solution, the reduced HTC interaction matrix, and the
cluster-expansion hierarchy, closure conditions, and numerical implementation
details.

\parhead{Notes}
The authors declare no competing financial interest.

\parhead{Acknowledgments}
This work was supported by the European Research Council under
ERC-2024-SyG-101167294; UnMySt.

\parhead{Data availability}
The data that support the findings of this work are available in \url{https://github.com/Okita0512/Collective-Polariton-Dynamics-Regimes}.

\bibliography{ref}

\newpage

\begin{figure}[t]
  \centering
  \includegraphics[width=1.0\columnwidth]{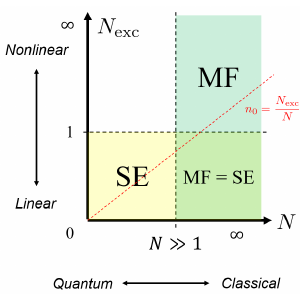}
  \caption{TOC Graphic}
  \label{fig:toc}
\end{figure}

\end{document}


\title[]{\large Supporting Information for: \\
Excitation density controlled regimes of collective light--matter dynamics}

\author{Wenxiang Ying}
\email{wying3@sas.upenn.edu}
\affiliation{Department of Chemistry, University of Pennsylvania, Philadelphia, Pennsylvania 19104, USA}

\author{Abraham Nitzan}
\email{anitzan@sas.upenn.edu}
\affiliation{Department of Chemistry, University of Pennsylvania, Philadelphia, Pennsylvania 19104, USA}
\affiliation{School of Chemistry, Tel Aviv University, Tel Aviv 69978, Israel}

\maketitle
\tableofcontents

\newpage
\section{Mean-field equations of motion}
\label{sec:SI_MF}

This section derives the mean-field (MF) equations used in the main text for the
TC and HTC models and fixes the notation used in the later dynamical reductions.

The MF approximation assumes a product factorization of the total density matrix,
\begin{align}
  \hat{\rho}(t) \approx \hat{\rho}_a(t)\otimes\bigotimes_{n=1}^{N}\hat{\rho}_n(t),
  \label{eq:SI_ansatz}
\end{align}
where $\hat{\rho}_a$ is the cavity reduced density matrix (RDM) and $\hat{\rho}_n$ is the
$n$-th emitter RDM.
Here, we specialize to the coherent-state
MF approximation~\cite{Keeling_PRL2022}. 
Eq.~\ref{eq:SI_ansatz} is the density-matrix generalization of the
time-dependent Hartree (TDH) product-state ansatz used in Ref.~\citenum{Cui_JCP2023}.
In the dissipation-free case, the Liouville--von Neumann (LvN) equation
$\dot{\hat{\rho}}=-i[\hat{H},\hat{\rho}]$ preserves the purity of the total state;
if the initial state is pure, Eq.~\ref{eq:SI_ansatz} is then equivalent to
the TDH wavefunction factorization
\begin{align}
    |\Psi(t)\rangle = |\alpha(t)\rangle\prod_{n=1}^{N}|\psi_n(t)\rangle,
\end{align}
where $|\alpha(t)\rangle$ is a coherent state of the cavity field
($\hat{a}|\alpha\rangle=\alpha|\alpha\rangle$) and $|\psi_n(t)\rangle$ is the
state of the $n$-th molecule, so that $\hat{\rho}_a=|\alpha\rangle\langle\alpha|$
and $\hat{\rho}_n=|\psi_n\rangle\langle\psi_n|$.
The density-matrix form in Eq.~\ref{eq:SI_ansatz} is the more general starting point.
When used below, cavity loss is included phenomenologically at the amplitude level
through the non-Hermitian substitution
$\hat{H}\to\hat{H}-i(\kappa/2)\hat{a}^\dagger\hat{a}$ in the cavity sector; this is
not a full Lindblad derivation.
We set $\hbar=1$ throughout and derive the MF equations of motion (EOM) for two cases.

\subsection{Case 1: Without vibration (TC model)}
\label{subsec:SI_TC_case1}

The TC Hamiltonian reads
\begin{align}
  \hat{H} &= \omega_\mathrm{c}\hat{a}^\dagger\hat{a}
            + \sum_{n=1}^{N}\omega_0\hat{\sigma}_n^+\hat{\sigma}_n^-
            + g_\mathrm{c}\sum_{n=1}^{N}
              \!\left(\hat{a}^\dagger\hat{\sigma}_n^- + \hat{a}\hat{\sigma}_n^+\right).
  \label{eq:SI_TC}
\end{align}
Applying the product ansatz of Eq.~\ref{eq:SI_ansatz} and taking partial traces of the
LvN equation $\dot{\hat\rho}=-i[\hat H,\hat\rho]$ yields equations
of motion for the emitter and cavity degrees of freedom (DOF).

Tracing over the cavity DOF gives the emitter EOM for molecule $n$,
\begin{align}
  \frac{\partial}{\partial t}\hat{\rho}_n
  = -i\bigl[\hat{H}_n^\mathrm{eff}(\alpha),\,\hat{\rho}_n\bigr],
  \label{eq:SI_rho_n}
\end{align}
with the effective Hamiltonian
\begin{align}
  \hat{H}_n^\mathrm{eff}(\alpha)
  = \omega_0\hat{\sigma}_n^+\hat{\sigma}_n^-
  + g_\mathrm{c}\!\left(\alpha^*\hat{\sigma}_n^- + \alpha\hat{\sigma}_n^+\right),
  \label{eq:SI_Heff}
\end{align}
where $\alpha\equiv\langle\hat{a}\rangle$ enters as a self-consistent mean field amplitude,
consistent with the time-dependent Hartree ansatz~\cite{Cui_JCP2022}.
Invoking permutation symmetry, all emitters share a single representative
RDM $\hat{\rho}_\mathrm{M}(t)$, so the emitter EOM becomes
\begin{align}
  \frac{\partial}{\partial t}\hat{\rho}_\mathrm{M}(t)
  = -i\bigl[\hat{H}^\mathrm{eff}(\alpha),\,\hat{\rho}_\mathrm{M}(t)\bigr],
  \label{eq:SI_rho_M}
\end{align}
with
\begin{align}
  \hat{H}^\mathrm{eff}(\alpha)
  = \omega_0\hat{\sigma}^+\hat{\sigma}^-
  + g_\mathrm{c}\!\left(\alpha^*\hat{\sigma}^- + \alpha\hat{\sigma}^+\right).
  \label{eq:SI_Heff_TC}
\end{align}

For the cavity field, the LvN equation under the ansatz gives
\begin{align}
  \frac{\partial}{\partial t}\hat{\rho}_a
  = -i\omega_\mathrm{c}[\hat{a}^\dagger\hat{a},\hat{\rho}_a]
    -ig_\mathrm{c}\sum_{n=1}^{N}
      \Bigl([\hat{a}^\dagger,\hat{\rho}_a]\,\sigma_n
           +[\hat{a},\hat{\rho}_a]\,\sigma_n^*\Bigr),
  \label{eq:SI_rho_a}
\end{align}
where $\sigma_n\equiv\mathrm{Tr}_n[\hat{\rho}_n\hat{\sigma}_n^-]$.
Multiplying by $\hat{a}$ and taking the trace
(using $\mathrm{Tr}[\hat{a}[\hat{a}^\dagger,\hat\rho_a]]=1$ and
$\mathrm{Tr}[\hat{a}[\hat{a},\hat\rho_a]]=0$) gives
\begin{align}
  \dot{\alpha} = -i\omega_\mathrm{c}\,\alpha - ig_\mathrm{c}\sum_{n=1}^{N}\sigma_n.
  \label{eq:SI_alpha_1}
\end{align}
Similarly, invoking permutation symmetry ($\sigma_n\to\sigma$) and adding cavity
loss $\kappa$, Eq.~\ref{eq:SI_alpha_1} becomes
\begin{align}
  \dot{\alpha}(t)
  = -i\!\left(\omega_\mathrm{c} - i\frac{\kappa}{2}\right)\!\alpha(t)
    - ig_\mathrm{c} N\,\sigma(t),
  \label{eq:SI_alpha}
\end{align}
where $\sigma(t)\equiv\mathrm{Tr}[\hat{\rho}_\mathrm{M}(t)\,\hat{\sigma}^-]$.
Eqs.~\ref{eq:SI_rho_M}--\ref{eq:SI_alpha} are the MF EOM for the TC model.

Equivalently, the same result follows from the Heisenberg equations under the MF
decoupling
\begin{equation}
\langle \hat a\,\hat O_n\rangle \approx \langle \hat a\rangle \langle \hat O_n\rangle,
\qquad
\langle \hat a^\dagger \hat O_n\rangle \approx \langle \hat a^\dagger\rangle \langle \hat O_n\rangle,
\label{eq:MF_factorization}
\end{equation}
with MF variables
\begin{equation}
\alpha \equiv \langle \hat a\rangle,
\qquad
\sigma \equiv \frac{1}{N}\sum_{n=1}^N \langle \hat\sigma_n^- \rangle,
\qquad
w \equiv \frac{1}{N}\sum_{n=1}^N \langle \hat\sigma_n^z \rangle.
\label{eq:MF_variables}
\end{equation}
Using $[\hat a^\dagger \hat a,\hat a]=-\hat a$ and $[\hat\sigma_n^z,\hat\sigma_n^\pm]=\pm2\hat\sigma_n^\pm$,
\begin{align}
\dot{\hat a}
&= -i\omega_\mathrm{c} \hat a - i g_\mathrm{c} \sum_{n=1}^N \hat\sigma_n^-,
\label{eq:a_heis}\\
\dot{\hat\sigma}_n^-
&= -i\omega_0 \hat\sigma_n^- + i g_\mathrm{c} \hat a\,\hat\sigma_n^z,
\label{eq:sigma_heis}\\
\dot{\hat\sigma}_n^z
&= 2i g_\mathrm{c}\left(\hat a^\dagger \hat\sigma_n^- - \hat a\,\hat\sigma_n^+\right).
\label{eq:w_heis}
\end{align}
Taking expectation values and applying Eq.~\ref{eq:MF_factorization} yields
\begin{align}
\dot{\alpha}
&= -i\omega_\mathrm{c} \alpha - i g_\mathrm{c} N\sigma,
\label{eq:alpha_full}\\
\dot{\sigma}
&= -i\omega_0 \sigma + i g_\mathrm{c} \alpha\, w,
\label{eq:sigma_full}\\
\dot{w}
&= 2i g_\mathrm{c}(\alpha^*\sigma-\alpha\sigma^*).
\label{eq:w_full}
\end{align}
In the resonant rotating frame ($\alpha \to \alpha e^{i\omega_\mathrm{c}t}$ and $\sigma \to \sigma e^{i\omega_0 t}$, with $\omega_0=\omega_\mathrm{c}$) and the lossless condition ($\kappa=0$), these reduce to
the Maxwell--Bloch equations the same as Eqs.~2 of the main text (differs only by a rotating frame),
\begin{equation}
\boxed{
\dot{\alpha}=-i g_\mathrm{c} N\sigma,\qquad
\dot{\sigma}= i g_\mathrm{c} \alpha w,\qquad
\dot{w}=2i g_\mathrm{c}(\alpha^*\sigma-\alpha\sigma^*).
}
\label{eq:MF_MB_rot}
\end{equation}
The Bloch-vector constraint $w^2+4|\sigma|^2\leq 1$ and the conserved excitation
number
$N_\mathrm{exc}=|\alpha|^2+\dfrac{N}{2}(1+w)$ are preserved by the lossless MF
equations.
Unlike the standard externally driven Bloch equations (where the field is prescribed),
Eqs.~\ref{eq:MF_MB_rot} are self-consistent: $\alpha(t)$ is a dynamical variable
whose back-action on the matter equations produces amplitude-dependent (anharmonic) Rabi
oscillations and an effective Duffing equation for $\alpha(t)$.

We now spell out the reduction used for Eq.~3 of the main text.  In the
lossless resonant rotating frame, the conserved excitation number is
\begin{align}
  N_\mathrm{exc}
  &= |\alpha|^2 + \frac{N}{2}(1+w).
  \label{eq:SI_Nexc_MF}
\end{align}
Its conservation follows directly from Eq.~\ref{eq:MF_MB_rot}:
\begin{align}
  \frac{d}{dt}|\alpha|^2
  &= \alpha^*\dot{\alpha}+\alpha\dot{\alpha}^*
   = -ig_\mathrm{c}N\alpha^*\sigma+ig_\mathrm{c}N\alpha\sigma^*,\\
  \frac{N}{2}\dot{w}
  &= ig_\mathrm{c}N(\alpha^*\sigma-\alpha\sigma^*),
\end{align}
so the two contributions cancel in $dN_\mathrm{exc}/dt$.  For the initial
condition used in the main text, $\alpha(0)=\alpha_0$, $\sigma(0)=0$, and
$w(0)=-1$, one has $N_\mathrm{exc}=|\alpha_0|^2$.  Defining the excitation
density $n_0=|\alpha_0|^2/N$, Eq.~\ref{eq:SI_Nexc_MF} gives
\begin{align}
  w(t)
  &= -1 + \frac{2}{N}\!\left(N_\mathrm{exc}-|\alpha(t)|^2\right)
   = -1 + 2n_0 - \frac{2|\alpha(t)|^2}{N}.
  \label{eq:SI_w_from_Nexc}
\end{align}

The Duffing equation then follows by eliminating $\sigma$ and $w$ from the first
Maxwell--Bloch equation, $\dot{\alpha}=-i g_\mathrm{c} N\sigma$.  Differentiating
it and using
$\dot{\sigma}=ig_\mathrm{c}\alpha w$ gives
\begin{align}
  \ddot{\alpha}
  &= -ig_\mathrm{c}N\dot{\sigma}
   = g_\mathrm{c}^2N\alpha w.
  \label{eq:SI_alpha_second_derivative}
\end{align}
Substituting Eq.~\ref{eq:SI_w_from_Nexc} into
Eq.~\ref{eq:SI_alpha_second_derivative} yields
\begin{align}
  \ddot{\alpha}
  + g_\mathrm{c}^2N(1-2n_0)\alpha
  + 2g_\mathrm{c}^2|\alpha|^2\alpha
  = 0.
  \label{eq:SI_complex_duffing}
\end{align}
For the phase convention $\alpha_0\in\mathbb{R}$, the rotating-frame solution has
$\alpha(t)\in\mathbb{R}$ and Eq.~\ref{eq:SI_complex_duffing} reduces to the
main-text Duffing equation,
\begin{align}
  \ddot{\alpha}
  + g_\mathrm{c}^2N(1-2n_0)\alpha
  + 2g_\mathrm{c}^2\alpha^3
  = 0.
  \label{eq:SI_Duffing_from_MBE}
\end{align}

\subsection{Case 2: With vibration (HTC model)}
\label{subsec:SI_HTC_case2}

The derivation follows Sec.~\ref{subsec:SI_TC_case1} identically; the only change is the
molecular Hamiltonian, which now includes a vibrational mode coupled to the electronic
excitation:
\begin{align}
  \hat{H}_n &= \omega_0\hat{\sigma}_n^+\hat{\sigma}_n^-
              + \nu\hat{b}_n^\dagger\hat{b}_n
              + \hat{\sigma}_n^+\hat{\sigma}_n^-\otimes c_\nu(\hat{b}_n+\hat{b}_n^\dagger),
  \label{eq:SI_Hn_vib}
\end{align}
where $\hat{b}_n$ ($\hat{b}_n^\dagger$) annihilates (creates) a vibrational quantum of
frequency $\nu$ and $c_\nu$ is the Huang--Rhys coupling strength.
The emitter EOM takes the same form as Eq.~\ref{eq:SI_rho_M},
\begin{align}
  \frac{\partial}{\partial t}\hat{\rho}_\mathrm{M}(t)
  = -i\bigl[\hat{H}^\mathrm{eff}(\alpha),\,\hat{\rho}_\mathrm{M}(t)\bigr],
  \label{eq:SI_rho_M_HTC}
\end{align}
but with the $\alpha$-dependent effective Hamiltonian, i.e., the single-molecule
form of the HTC Hamiltonian used in the main text,
\begin{align}
  \hat{H}^\mathrm{eff}(\alpha)
  &= \omega_0|e\rangle\langle e|
     + \nu\hat{b}^\dagger\hat{b}
     + c_\nu|e\rangle\langle e|\otimes(\hat{b}+\hat{b}^\dagger)
     + g_\mathrm{c}\!\left(\alpha^*|g\rangle\langle e| + \alpha|e\rangle\langle g|\right),
  \label{eq:SI_Heff_vib}
\end{align}
where $\{|g\rangle,|e\rangle\}$ are the electronic ground and excited states and
permutation symmetry $\hat{\rho}_n\to\hat{\rho}_\mathrm{M}$,
$\hat{b}_n\to\hat{b}$ has been applied.
The cavity EOM~\ref{eq:SI_alpha} is unchanged; $\sigma(t)=\mathrm{Tr}[\hat{\rho}_\mathrm{M}(t)\,\hat{\sigma}^-]$
now denotes the electronic coherence traced over both electronic and vibrational DOF.
Eqs.~\ref{eq:SI_rho_M_HTC} and~\ref{eq:SI_alpha} with $\hat{H}^\mathrm{eff}$
of Eq.~\ref{eq:SI_Heff_vib} constitute the full MF EOM for the HTC model.
To be explicit, the complete MF EOM for the HTC model are:
\label{eq:SI_MF_HTC_full}
\begin{align}
  \frac{\partial}{\partial t}\hat{\rho}_\mathrm{M}(t)
  &= -i\bigl[\hat{H}^\mathrm{eff}(\alpha),\,\hat{\rho}_\mathrm{M}(t)\bigr],
  \label{eq:SI_MF_HTC_matter}\\
  \dot{\alpha}(t)
  &= -i\!\left(\omega_\mathrm{c}-i\frac{\kappa}{2}\right)\!\alpha(t)
     - ig_\mathrm{c}N\,\sigma(t),
  \label{eq:SI_MF_HTC_cavity}
\end{align}
where $\hat{H}^\mathrm{eff}(\alpha)$ is given by Eq.~\ref{eq:SI_Heff_vib} and
$\sigma(t)\equiv\mathrm{Tr}[\hat{\rho}_\mathrm{M}(t)\,\hat{\sigma}^-]$ is the
electronic coherence traced over both electronic and vibrational DOF.
The cavity amplitude $\alpha(t)$ drives the single-molecule RDM through the effective
Hamiltonian and, in turn, receives its back-action through $\sigma(t)$, making
Eqs.~\ref{eq:SI_MF_HTC_full} self-consistent.

In particular, when the vibrational Fock space is truncated to two states
$\{|a\rangle,|b\rangle\}$ (vibrational ground and first excited level), the
single-molecule MF state takes the explicit form
\begin{align}
|\psi\rangle = c_{ga}|g,a\rangle + c_{ea}|e,a\rangle
             + c_{gb}|g,b\rangle + c_{eb}|e,b\rangle,
\label{eq:SI_psi_2fock}
\end{align}
where $\{|g\rangle,|e\rangle\}$ are the electronic states and $\{|a\rangle,|b\rangle\}$
are the two vibrational Fock states.
The electronic coherence is obtained by tracing over the vibrational DOF,
\begin{align}
\sigma(t) \equiv \mathrm{Tr}[\hat{\rho}_\mathrm{M}(t)\,\hat{\sigma}^-]
= \sum_{\mu\in\{a,b\}}\langle g,\mu|\hat{\rho}_\mathrm{M}(t)|e,\mu\rangle
= c_{ga}^*(t)\,c_{ea}(t) + c_{gb}^*(t)\,c_{eb}(t).
\label{eq:SI_sigma_2fock}
\end{align}
Substituting Eq.~\ref{eq:SI_sigma_2fock} into Eq.~\ref{eq:SI_MF_HTC_cavity}
and setting $\kappa=0$ gives
\begin{equation}
\dot{\alpha} = -i\omega_\mathrm{c}\,\alpha
  - ig_\mathrm{c}N\bigl(c_{ga}^*c_{ea} + c_{gb}^*c_{eb}\bigr),
\label{eq:SI_alpha_2fock}
\end{equation}
which is the two-vibrational-state reduction of Eq.~\ref{eq:SI_MF_HTC_cavity}.

The single-molecule Schr\"{o}dinger equation $i\partial_t|\psi\rangle = \hat{H}^\mathrm{eff}(\alpha)|\psi\rangle$, with the ansatz in Eq.~\ref{eq:SI_psi_2fock} and Hamiltonian in Eq.~\ref{eq:SI_Heff_vib}, yields one amplitude equation per basis state.
Restricting $\hat{b}+\hat{b}^\dagger$ to the vibrational subspace $\{|a\rangle,|b\rangle\}$, the only nonzero off-diagonal matrix element is
\begin{equation}
  \langle a|\hat{b}+\hat{b}^\dagger|b\rangle = 1,
  \label{eq:SI_bpbdag_trunc}
\end{equation}
so projecting onto $\{|g,a\rangle,|e,a\rangle,|g,b\rangle,|e,b\rangle\}$ gives the amplitude equations
\label{eq:SI_amp_2fock}
\begin{align}
  \dot{c}_{ga} &= -ig_\mathrm{c}\alpha^*\,c_{ea},
  \label{eq:SI_cga}\\
  \dot{c}_{ea} &= -i\omega_0\,c_{ea} - ic_\nu\,c_{eb} - ig_\mathrm{c}\alpha\,c_{ga},
  \label{eq:SI_cea}\\
  \dot{c}_{gb} &= -i\nu\,c_{gb} - ig_\mathrm{c}\alpha^*\,c_{eb},
  \label{eq:SI_cgb}\\
  \dot{c}_{eb} &= -i(\omega_0+\nu)\,c_{eb} - ic_\nu\,c_{ea} - ig_\mathrm{c}\alpha\,c_{gb}.
  \label{eq:SI_ceb}
\end{align}
These are the single-molecule TDH equations implied by the MF EOM in Eq.~\ref{eq:SI_MF_HTC_full}.
Together with Eq.~\ref{eq:SI_alpha_2fock}, they form a closed, self-consistent set of five coupled equations for the two-vibrational-state HTC model.

\newpage
\section{Duffing dynamics and amplitude-dependent Rabi frequency}
\label{sec:SI_Duffing}

This section supports the main-text statement that finite excitation density
$n_0$ produces nonlinear MF dynamics rather than a failure of the MF approximation.

In the lossless, resonant MF setting, the collective cavity amplitude
obeys the Duffing equation
\begin{equation}
\ddot{\alpha} + \omega_0^2 \alpha + \beta \alpha^3 = 0,
\label{eq:duffing_general}
\end{equation}
where, from the main-text Duffing reduction,
\begin{equation}
\omega_0^2 = g_\mathrm{c}^2 N (1 - 2 n_0),
\qquad
\beta = 2 g_\mathrm{c}^2.
\end{equation}
Here $n_0 = \alpha_0^2/N$ denotes the initial excitation density.
We consider the initial conditions
\begin{equation}
\alpha(0) = A \equiv \alpha_0,
\qquad
\dot{\alpha}(0) = 0.
\end{equation}

\subsection{Energy integral and exact reduction}
\label{subsec:energy_integral}

Eq.~\ref{eq:duffing_general} is conservative and admits the
energy integral
\begin{equation}
E = \frac{1}{2}\dot{\alpha}^2
+ \frac{1}{2}\omega_0^2 \alpha^2
+ \frac{1}{4}\beta \alpha^4.
\end{equation}
With the chosen initial conditions,
\begin{equation}
E = \frac{1}{2}\omega_0^2 A^2 + \frac{1}{4}\beta A^4.
\end{equation}
Hence,
\begin{equation}
\dot{\alpha}^2
=
\omega_0^2 (A^2 - \alpha^2)
+ \frac{\beta}{2} (A^4 - \alpha^4).
\label{eq:alpha_energy}
\end{equation}
Factoring the right-hand side,
\begin{equation}
\dot{\alpha}^2
=
(A^2 - \alpha^2)
\left[
\omega_0^2
+ \frac{\beta}{2}(A^2 + \alpha^2)
\right].
\label{eq:SI_factored}
\end{equation}

\subsection{Solution via separation of variables and elliptic integrals}
\label{subsec:elliptic_integrals}

Starting from Eq.~\ref{eq:SI_factored},
\begin{equation}
\dot{\alpha}^2
=
(A^2-\alpha^2)
\left[
\omega_0^2 + \frac{\beta}{2}(A^2+\alpha^2)
\right],
\label{eq:S_sep_start}
\end{equation}
we now solve the equation by separation of variables.
Choosing the branch with $\dot{\alpha}\le 0$ immediately after $t=0$,
\begin{equation}
\dot{\alpha}
=
-
\sqrt{
(A^2-\alpha^2)
\left[
\omega_0^2 + \frac{\beta}{2}(A^2+\alpha^2)
\right]
}.
\end{equation}
Separating variables gives
\begin{equation}
dt
=
-\frac{d\alpha}{
\sqrt{
(A^2-\alpha^2)
\left[
\omega_0^2 + \frac{\beta}{2}(A^2+\alpha^2)
\right]
}
}.
\label{eq:S_sep1}
\end{equation}

\paragraph*{Dimensionless scaling.}
Introduce the dimensionless variable
\begin{equation}
x \equiv \frac{\alpha}{A},
\qquad x(0)=1,
\end{equation}
so that $\alpha = A x$ and $d\alpha = A\,dx$.
Eq.~\ref{eq:S_sep1} becomes
\begin{equation}
dt
=
-\frac{dx}{
\sqrt{
(1-x^2)
\left[
\omega_0^2 + \frac{\beta A^2}{2}(1+x^2)
\right]
}
}.
\end{equation}

Define the auxiliary frequency scale
\begin{equation}
\omega_*^2 \equiv \omega_0^2 + \beta A^2,
\qquad
m \equiv \frac{\beta A^2}{2(\omega_0^2 + \beta A^2)}
= \frac{\beta A^2}{2\omega_*^2}.
\label{eq:S_omega_star_m}
\end{equation}
A short algebraic manipulation shows
\begin{equation}
\omega_0^2 + \frac{\beta A^2}{2}(1+x^2)
=
\omega_*^2 \left[ 1 - m(1-x^2) \right].
\end{equation}
Hence
\begin{equation}
dt
=
-\frac{dx}{
\omega_*
\sqrt{(1-x^2)\left[1-m(1-x^2)\right]}
}.
\label{eq:S_sep2}
\end{equation}

\paragraph*{Reduction to the canonical elliptic integral.}
Let
\begin{equation}
x = \cos\theta,
\qquad
\theta(0)=0.
\end{equation}
Then $dx=-\sin\theta\,d\theta$ and $1-x^2=\sin^2\theta$.
Substituting into Eq.~\ref{eq:S_sep2} gives
\begin{equation}
dt
=
\frac{d\theta}{
\omega_*\sqrt{1-m\sin^2\theta}
}.
\end{equation}
Integrating from $0$ to $t$ yields
\begin{equation}
\omega_* t
=
\int_0^{\theta(t)}
\frac{d\phi}{\sqrt{1-m\sin^2\phi}}
\equiv
F(\theta(t)\mid m),
\label{eq:S_incomplete}
\end{equation}
where
\begin{equation}
F(\theta\mid m)
=
\int_0^{\theta}
\frac{d\phi}{\sqrt{1-m\sin^2\phi}}
\end{equation}
is the incomplete elliptic integral of the first kind.

\paragraph*{Inversion and Jacobi elliptic functions.}
Eq.~\ref{eq:S_incomplete} defines $\theta$ implicitly as a function of $t$.
Its inverse is written using the Jacobi amplitude function,
\begin{equation}
\theta(t) = \mathrm{am}(\omega_* t \mid m).
\end{equation}
The Jacobi elliptic cosine function is then defined by
\begin{equation}
\mathrm{cn}(u\mid m)
=
\cos\!\bigl(\mathrm{am}(u\mid m)\bigr).
\end{equation}
Since $x=\cos\theta$, we obtain
\begin{equation}
x(t)
=
\mathrm{cn}(\omega_* t \mid m),
\end{equation}
and therefore the exact solution
\begin{equation}
\boxed{
\alpha(t)
=
A\,\mathrm{cn}(\omega_* t \mid m).
}
\end{equation}

\paragraph*{Harmonic limit.}
In the limit $m\to 0$ (i.e., $\beta A^2 \ll \omega_0^2$),
\begin{equation}
F(\theta\mid 0)=\theta,
\qquad
\mathrm{am}(u\mid 0)=u,
\qquad
\mathrm{cn}(u\mid 0)=\cos u,
\end{equation}
so the solution smoothly reduces to the harmonic oscillation
$\alpha(t)=A\cos(\omega_0 t)$.

\subsection{Jacobi elliptic functions: definitions and properties}
\label{subsec:jacobi}

Only a few Jacobi-function identities are needed to verify the Duffing solution.
The Jacobi amplitude function $\mathrm{am}(u\mid m)$ is defined by
\begin{equation}
u = F(\theta\mid m)
= \int_0^{\theta}
\frac{d\phi}{\sqrt{1-m\sin^2\phi}},
\end{equation}
and the functions used below are
\begin{equation}
\mathrm{sn}(u\mid m) = \sin(\mathrm{am}(u\mid m)),
\qquad
\mathrm{cn}(u\mid m) = \cos(\mathrm{am}(u\mid m)),
\qquad
\mathrm{dn}(u\mid m)
=
\sqrt{1-m\,\mathrm{sn}^2(u\mid m)}.
\end{equation}
In the harmonic limit,
\begin{equation}
\mathrm{sn}(u\mid 0)=\sin u,
\qquad
\mathrm{cn}(u\mid 0)=\cos u,
\qquad
\mathrm{dn}(u\mid 0)=1.
\end{equation}
The differential identity needed below is
\begin{equation}
\boxed{
\frac{d^2}{du^2}\mathrm{cn}(u\mid m)
=
(2m-1)\,\mathrm{cn}(u\mid m)
-2m\,\mathrm{cn}^3(u\mid m),
}
\label{eq:cn_second_order}
\end{equation}
Using Eq.~\ref{eq:cn_second_order} with
\begin{equation}
\alpha(t)=A\,\mathrm{cn}(\omega_* t\mid m).
\end{equation}
Then
\begin{equation}
\ddot{\alpha}
=
A\omega_*^2
\left[
(2m-1)\mathrm{cn}(\omega_* t\mid m)
-2m\,\mathrm{cn}^3(\omega_* t\mid m)
\right].
\end{equation}
Identifying $\alpha=A\,\mathrm{cn}$ and $\alpha^3=A^3\mathrm{cn}^3$,
the Duffing equation
\begin{equation}
\ddot{\alpha}+\omega_0^2\alpha+\beta\alpha^3=0
\end{equation}
is satisfied exactly provided
\begin{equation}
\omega_*^2=\omega_0^2+\beta A^2,
\qquad
m=\frac{\beta A^2}{2(\omega_0^2+\beta A^2)}.
\end{equation}
Hence the parametrization $\alpha(t)=A\,\mathrm{cn}(\omega_* t\mid m)$
is not an ansatz but the solution generated by inversion of the
elliptic integral arising from energy conservation.

\subsection{Exact oscillation period}
\label{subsec:exact_period}

The real period of $\mathrm{cn}(u\mid m)$ is $4K(m)$,
where
\begin{equation}
K(m)
=
\int_0^{\pi/2}
\frac{d\theta}{\sqrt{1-m\sin^2\theta}}
\end{equation}
is the complete elliptic integral of the first kind.
Therefore, the exact oscillation period is
\begin{equation}
T = \frac{4K(m)}{\omega_*},
\end{equation}
and the exact effective Rabi frequency, consistent with the notation in the
main text, is
\begin{equation}
\boxed{
\Omega^\mathrm{eff}
=
\frac{4\pi}{T}
=
\frac{\pi\,\omega_*}{K(m)}.
}
\end{equation}

\subsection{Small-amplitude expansion of the effective Rabi frequency}
\label{subsec:small_amplitude}

For weak nonlinearity ($m \ll 1$), the elliptic integral admits
\begin{equation}
K(m)
=
\frac{\pi}{2}
\left(
1 + \frac{m}{4}
+ \frac{9m^2}{64}
+ O(m^3)
\right).
\end{equation}
Hence,
\begin{equation}
\Omega^\mathrm{eff}
=
2\omega_*
\left(
1
- \frac{m}{4}
- \frac{5m^2}{64}
+ O(m^3)
\right).
\end{equation}
Expanding in the small parameter
$\epsilon = \beta A^2 / \omega_0^2$ yields
\begin{equation}
\boxed{
\Omega^\mathrm{eff}
=
2\omega_0
\left[
1
+
\frac{3}{8}\frac{\beta A^2}{\omega_0^2}
-
\frac{21}{256}
\left(
\frac{\beta A^2}{\omega_0^2}
\right)^2
+
O\!\left(
\frac{\beta A^2}{\omega_0^2}
\right)^3
\right].
}
\label{eq:freq_shift_general}
\end{equation}

\subsection{Application to the MF polariton problem}
\label{subsec:MF_polariton_app}

For the MF Duffing equation derived in the main text,
\begin{equation}
\beta = 2g_\mathrm{c}^2,
\qquad
\omega_0^2 = g_\mathrm{c}^2 N (1 - 2 n_0),
\qquad
A^2 = \alpha_0^2 = n_0 N.
\end{equation}
Thus,
\begin{equation}
\frac{\beta A^2}{\omega_0^2}
=
\frac{2n_0}{1-2n_0},
\end{equation}
and Eq.~\ref{eq:freq_shift_general} gives
\begin{equation}
\boxed{
\Omega^\mathrm{eff}
=
2g_\mathrm{c}\sqrt{N(1-2n_0)}
\left[
1
+
\frac{3n_0}{4(1-2n_0)}
-
\frac{21 n_0^2}{64(1-2n_0)^2}
+
\cdots
\right].
}
\end{equation}

Equation above makes explicit how increasing excitation density
($n_0$ increasing, or equivalently $w$ deviating from $-1$)
induces an amplitude-dependent renormalization of the Rabi oscillation
frequency. In the present hardening Duffing case ($\beta>0$),
the oscillation frequency increases with amplitude; if the cubic term
had opposite sign, the same procedure would lead to frequency softening.
To leading order in $n_0\ll 1$, this reduces to
$\Omega^\mathrm{eff}\approx 2g_\mathrm{c}\sqrt{N}(1-n_0/4)$, matching the
main-text notation.

\paragraph*{Direct harmonic-balance derivation.}
The leading anharmonic frequency correction can be obtained more directly---without
invoking elliptic integrals---via a harmonic-balance (secular-perturbation) argument.
Substitute the single-frequency trial solution
\begin{equation}
\alpha(t) = \alpha_0\cos(\tilde\omega\, t)
\label{eq:hb_ansatz}
\end{equation}
into the Duffing equation in Eq.~\ref{eq:duffing_general}.
Using $\ddot\alpha = -\alpha_0\tilde\omega^2\cos(\tilde\omega t)$ and the identity
\begin{equation}
\cos^3(\tilde\omega t) = \dfrac{3}{4}\cos(\tilde\omega t)+\dfrac{1}{4}\cos(3\tilde\omega t),
\end{equation}
one obtains
\begin{equation}
\left(-\tilde\omega^2+\omega_0^2+\dfrac{3}{4}\beta\alpha_0^2\right)
\alpha_0\cos(\tilde\omega t)
+\dfrac{1}{4}\beta\alpha_0^3\cos(3\tilde\omega t)=0.
\end{equation}
Neglecting the third-harmonic term (which is small when $n_0\ll 1$) and demanding
that the coefficient of the secular $\cos(\tilde\omega t)$ component vanishes gives the
renormalized oscillation frequency
\begin{equation}
\tilde\omega^2 = \omega_0^2 + \dfrac{3}{4}\beta\alpha_0^2.
\label{eq:hb_freq}
\end{equation}

Inserting the polariton parameters $\omega_0^2 = g_\mathrm{c}^2 N(1-2n_0)$,
$\beta = 2g_\mathrm{c}^2$, and $\alpha_0^2 = Nn_0$ into Eq.~\ref{eq:hb_freq}:
\begin{align}
\tilde\omega^2
&= g_\mathrm{c}^2 N(1-2n_0) + \dfrac{3}{4}(2g_\mathrm{c}^2)(Nn_0) \notag\\
&= g_\mathrm{c}^2 N\!\left(1-2n_0+\dfrac{3n_0}{2}\right)
 = g_\mathrm{c}^2 N\!\left(1-\dfrac{n_0}{2}\right).
\label{eq:hb_omega_sq}
\end{align}
Taking the square root and expanding for $n_0\ll 1$:
\begin{equation}
\tilde\omega
= g_\mathrm{c}\sqrt{N}\sqrt{1-\dfrac{n_0}{2}}
\approx g_\mathrm{c}\sqrt{N}\!\left(1-\dfrac{n_0}{4}\right).
\label{eq:hb_omega_leading}
\end{equation}

\paragraph*{From oscillation frequency to Rabi frequency.}
The quantity $\tilde\omega$ is the oscillation frequency of the cavity amplitude
$\alpha(t)$.  The physical Rabi frequency is conventionally defined from the
population oscillation.  From the conservation law
$w = -1+2(n_0-|\alpha|^2/N)$, one has
\begin{equation}
w(t) = -1+2n_0\bigl[1-\cos^2(\tilde\omega t)\bigr] = -(1-n_0) - n_0\cos(2\tilde\omega t),
\end{equation}
so the excitation probability $P_\mathrm{ex}=(1+w)/2$ oscillates at frequency
$2\tilde\omega$.  Hence the effective Rabi frequency is
\begin{equation}
\Omega^\mathrm{eff} \equiv 2\tilde\omega
\approx 2g_\mathrm{c}\sqrt{N}\!\left(1-\frac{n_0}{4}\right),
\qquad n_0\ll 1,
\label{eq:Omega_eff_hb}
\end{equation}
in agreement with the main-text result and with the leading-order expansion of the exact
elliptic formula derived above.

A remark on the sign of the correction is in order.  The cubic coefficient
$\beta = 2g_\mathrm{c}^2>0$ corresponds to a \emph{hardening} Duffing oscillator:
the cubic term alone would \emph{increase} the oscillation frequency with amplitude.
The net downward shift seen in Eq.~\ref{eq:Omega_eff_hb} is instead dominated by the
$(1-2n_0)$ factor in $\omega_0^2$, which encodes the saturation of the two-level
inversion; the cubic (anharmonic) correction $+\dfrac{3n_0}{2}$ partially compensates
that suppression, so the leading-order correction to the bare Rabi frequency
$2g_\mathrm{c}\sqrt{N}$ is $-n_0/4$, smaller in magnitude than the $-n_0$ one would
obtain from $\omega_0$ alone.

\subsection{Fixed points and pitchfork bifurcation}
\label{subsec:duffing_fixed_points}

This subsection records an optional mathematical characterization of the conservative
reduced MF equation. Because the Duffing equation in Eq.~\ref{eq:duffing_general} is
conservative, its
fixed points are the time-independent solutions satisfying $\ddot\alpha=\dot\alpha=0$.
Setting $\ddot\alpha = 0$ gives
\begin{equation}
  \alpha_\mathrm{ss}\!\left[g_\mathrm{c}^2 N(1-2n_0)+2g_\mathrm{c}^2\alpha_\mathrm{ss}^2\right]=0.
\end{equation}
This admits a \emph{trivial} solution $\alpha_\mathrm{ss}=0$ and, when $n_0>1/2$,
two \emph{nontrivial} equilibria
\begin{equation}
  \alpha_\mathrm{ss}
  = \pm\sqrt{N\!\left(n_0-\dfrac{1}{2}\right)},
  \qquad n_0 > \dfrac{1}{2}.
  \label{eq:alpha_ss}
\end{equation}
These fixed points are more transparently understood through the effective potential
\begin{equation}
  V(\alpha)
  = \dfrac{1}{2}g_\mathrm{c}^2 N(1-2n_0)\,\alpha^2
  + {1}{2}g_\mathrm{c}^2\,\alpha^4,
  \label{eq:Veff}
\end{equation}
which takes the canonical Landau form $V = a\alpha^2+b\alpha^4$ with control
parameter $a\propto(1-2n_0)$ and $b=g_\mathrm{c}^2/2>0$.
For $n_0<1/2$ the coefficient $a>0$ and $V$ has a single minimum at
$\alpha=0$; for $n_0>1/2$ the coefficient $a$ changes sign, converting $V$
into a double-well potential with minima at Eq.~\ref{eq:alpha_ss}.
This has the algebraic form of a \emph{supercritical pitchfork bifurcation}:
the trivial state loses stability at $n_0=1/2$ and two new equilibria bifurcate
continuously from it, spontaneously breaking the $\alpha\to-\alpha$ symmetry.

Using the conservation law $w=-1+2(\alpha_0^2-\alpha^2)/N$, one can verify that
the nontrivial fixed points satisfy $w_\mathrm{ss}=0$ (zero inversion), which is
physically reasonable as a half-inversion state.

\paragraph*{Remark on the conservative nature.}
It is essential to note that Eq.~\ref{eq:duffing_general} contains no damping;
the fixed points above are therefore \emph{equilibrium points} of a Hamiltonian
flow, not dissipative steady states.
Unless initialized exactly at $(\alpha_\mathrm{ss},\dot\alpha=0)$, the system
simply oscillates around the relevant potential minimum.
Thus the bifurcation-like structure should be interpreted only as a property of
the conservative reduced MF equation. It is not a dissipative phase transition and
is not used as evidence for chaos in this work. In the present context it simply
marks the change from a single-well to a double-well effective potential as the
initial excitation density crosses $n_0=1/2$.

\newpage
\section{Cluster-expansion corrections to mean field}
\label{sec:SI_CE}

This section discusses cluster-expansion (CE) corrections for the resonant
Jaynes--Cummings/Tavis--Cummings problem without vibrations. 

We use the convention
$\hat\sigma_i^z=|e_i\rangle\langle e_i|-|g_i\rangle\langle g_i|$.  In the
lossless resonant rotating frame,
\begin{align}
  \hat H_\mathrm{TC}
  =
  g_\mathrm{c}\sum_{i=1}^{N}
  \left(\hat a^\dagger\hat\sigma_i^-+\hat a\hat\sigma_i^+\right),
\end{align}
so that
\begin{align}
  \dot{\hat a}=-ig_\mathrm{c}\sum_i\hat\sigma_i^-,
  \qquad
  \dot{\hat\sigma}_i^-=ig_\mathrm{c}\hat a\hat\sigma_i^z,
  \qquad
  \dot{\hat\sigma}_i^z
  =
  2ig_\mathrm{c}
  \left(\hat a^\dagger\hat\sigma_i^-
        -\hat a\hat\sigma_i^+\right).
\end{align}
Same-site Pauli matrix identities are always imposed before any cumulant closure, e.g.
\begin{align}
  \hat\sigma_i^+\hat\sigma_i^-=\frac{1+\hat\sigma_i^z}{2},\qquad
  \hat\sigma_i^-\hat\sigma_i^+=\frac{1-\hat\sigma_i^z}{2},\qquad
  \hat\sigma_i^-\hat\sigma_i^z=\hat\sigma_i^-.
\end{align}

\subsection{Mean-field closure}
\label{subsec:SI_CE_MF}

The MF approximation factorizes all molecule--field and molecule--molecule
moments into products of one-body averages.  Equivalently, all
correlations beyond first order are set to zero.  For any ordered product of
operators on distinct DOF, the MF contraction rule is
\begin{align}
  \langle \hat A\hat B\rangle_{\mathrm{MF}}
  =
  \langle \hat A\rangle\langle \hat B\rangle,
\end{align}
and similarly for higher products.  Same-site Pauli matrix identities are still applied
before this factorization.  For example,
\begin{align}
  \langle \hat a\hat\sigma_i^z\rangle_{\mathrm{MF}}
  =\alpha w,\qquad
  \langle \hat\sigma_i^+\hat\sigma_j^-\rangle_{\mathrm{MF}}
  =s^*s\quad (i\ne j).
\end{align}
Under permutation symmetry,
\begin{align} \label{eq:op_1-body}
  \alpha=\langle\hat a\rangle,\qquad
  s=\langle\hat\sigma_i^-\rangle,\qquad
  w=\langle\hat\sigma_i^z\rangle
\end{align}
are sufficient for any $N$.  The lossless resonant MF equations are (c.f. Eq.~\ref{eq:MF_MB_rot} and replace $\sigma$ by $s$)
\begin{align}
  \dot\alpha&=-ig_\mathrm{c}Ns,\\
  \dot s&=ig_\mathrm{c}\alpha w,\\
  \dot w&=2ig_\mathrm{c}\left(\alpha^*s-\alpha s^*\right).
  \label{eq:SI_CE_MF}
\end{align}
Thus the MF hierarchy has the same three dynamical variables for $N=1$ and
for arbitrary $N$; changing $N$ only changes the collective prefactor in the
cavity equation.  In the lossless case,
$N_\mathrm{exc}^{\mathrm{MF}}
  =
  |\alpha|^2+\dfrac{N}{2}(1+w)$
is conserved.

\subsection{\texorpdfstring{$N=1$}{N=1} cluster expansion}
\label{subsec:SI_N1_CE}

For the single-emitter benchmark the initial state is
$|1\rangle\otimes|g\rangle$.  The dynamics remains in
\begin{align}
  \mathcal H_{N_\mathrm{exc}=1}
  =
  \mathrm{span}\{|1,g\rangle,|0,e\rangle\},
\end{align}
where the identities
\begin{align}
  \hat n^2=\hat n,\qquad
  \hat n\hat a=0,\qquad
  \hat a^2=0,\qquad
  \hat n+\hat\sigma^+\hat\sigma^-=1
\end{align}
hold.  This identity-constrained setting is applied only to
single-emitter CE2--CE5 hierarchy.\\

{\bf CE2.} At CE2 we propagate the one- and two-body moments
\begin{align}
  y_{\mathrm{CE2}}^{(1)}
  =
  (\alpha,s,C,D,U,w,n),
\end{align}
with one-body moments $(\alpha, s, w)$ defined in Eq.~\ref{eq:op_1-body},
and two-body moments
\begin{align}
  n=\langle\hat a^\dagger\hat a\rangle,\qquad
  C=\langle\hat a\hat\sigma^+\rangle,\qquad
  D=\langle\hat a^\dagger\hat\sigma^-\rangle,\qquad
  U=\langle\hat a\hat\sigma^z\rangle .
\end{align}
The equations of motion are
\begin{align}
  \dot\alpha=-ig_\mathrm{c}s,\qquad
  \dot s=ig_\mathrm{c}U,\qquad
  \dot w=2ig_\mathrm{c}(D-C),\qquad
  \dot n=ig_\mathrm{c}(C-D),
\end{align}
which conserve $N_\mathrm{exc}=n+(1+w)/2$.
The remaining CE2 equations contain cubic moments,
\begin{align}
  \dot C
  &=
  -ig_\mathrm{c}
  \left[
    \langle\hat a\hat a^\dagger\hat\sigma^z\rangle
    +\frac{1-w}{2}
  \right],\\
  \dot D
  &=
  ig_\mathrm{c}
  \left[
    \langle\hat a^\dagger\hat a\hat\sigma^z\rangle
    +\frac{1+w}{2}
  \right],\\
  \dot U
  &=
  -ig_\mathrm{c}s
  +2ig_\mathrm{c}
  \left[
    \langle\hat a\hat a^\dagger\hat\sigma^-\rangle
    -\langle\hat a^2\hat\sigma^+\rangle
  \right].
\end{align}
CE2 closes these terms by setting third-order connected cumulants to zero,
\begin{align}
  \langle ABC\rangle_{\mathrm{CE2}}
  =
  \langle AB\rangle\langle C\rangle
  +\langle AC\rangle\langle B\rangle
  +\langle BC\rangle\langle A\rangle
  -2\langle A\rangle\langle B\rangle\langle C\rangle,
\end{align}
after the same-site identities above have been applied.\\

{\bf CE3.} CE3 promotes the cubic moments that enter the CE2 equations to independent
variables,
\begin{align}
  y_{\mathrm{CE3}}^{(1)}
  =
  (\alpha,s,C,D,U,S_1,Q,w,n,N_w),
\end{align}
where
\begin{align}
  S_1=\langle\hat a^\dagger\hat a\hat\sigma^-\rangle,\qquad
  Q=\langle\hat a^2\hat\sigma^+\rangle,\qquad
  N_w=\langle\hat a^\dagger\hat a\hat\sigma^z\rangle .
\end{align}
With these variables, the equations for $\dot C$, $\dot D$, and $\dot U$
are exact in terms of retained moments.  The new equations for
$\dot S_1$, $\dot Q$, and $\dot N_w$ generate quartic moments, which are
closed by setting fourth-order connected cumulants to zero.  For the
$|1\rangle\otimes|g\rangle$ initial state, the CE3 initial conditions are
$S_{1}(0)=Q(0)=0$ and $N_{w}(0)=-1$.\\

\begin{figure}[t]
  \includegraphics[width=0.7\columnwidth]{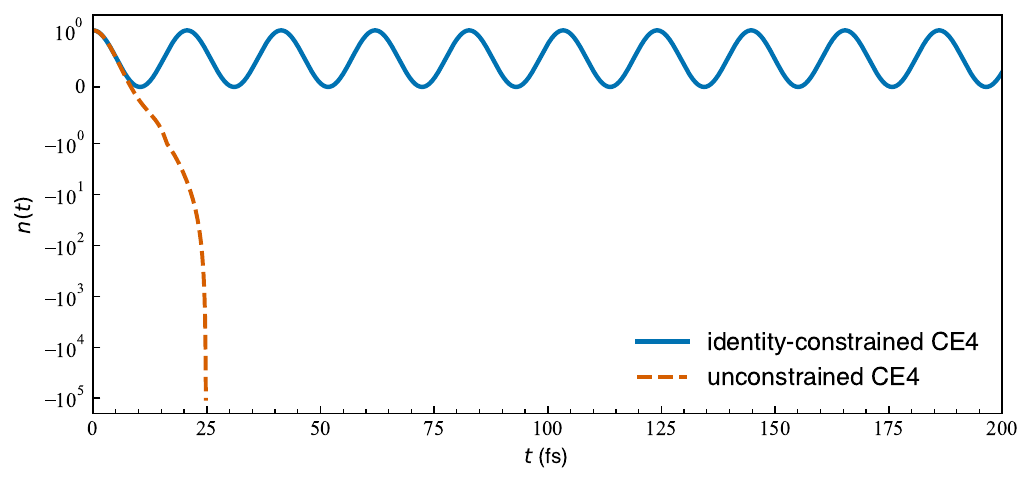}
  \caption{Role of the strict single-excitation operator identities in the
  $N=1$, $N_\mathrm{exc}=1$ CE4 benchmark.  The photon number obtained by
  imposing
  $T_1=\langle\hat n\hat a\hat\sigma^z\rangle=0$,
  $T_2=\langle\hat a\hat n\hat\sigma^z\rangle=U$,
  $C_n=\langle\hat n\hat a\hat\sigma^+\rangle=0$, and
  $S_d=\langle\hat n\hat a^\dagger\hat\sigma^-\rangle=D$
  remains bounded and follows the expected one-excitation Rabi dynamics.
  If these identities are not imposed, the promoted quartic moments leave
  the physical one-excitation manifold and the CE4 trajectory diverges
  rapidly.}
  \label{fig:SI_CE4_identity_constraints}
\end{figure}

{\bf CE4.} CE4 promotes the dominant quartic moments instead of factorizing them at CE3:
\begin{align}
  y_{\mathrm{CE4}}^{(1)}
  =
  (\alpha,s,C,D,U,S_1,Q,T_1,T_2,S_d,C_n,N_a,w,n,N_w,n_2),
\end{align}
with
\begin{align}
  T_1&=\langle\hat n\hat a\hat\sigma^z\rangle,\qquad
  T_2=\langle\hat a\hat n\hat\sigma^z\rangle,\qquad
  S_d=\langle\hat n\hat a^\dagger\hat\sigma^-\rangle,\qquad
  C_n=\langle\hat n\hat a\hat\sigma^+\rangle,\qquad
  N_a=\langle\hat n\hat a\rangle,\qquad
  n_2=\langle\hat n^2\rangle .
\end{align}
The fifth-order moments generated by these equations are closed by neglecting
the corresponding fifth-order connected cumulants.  In the strict
$N_\mathrm{exc}=1$ manifold, several quartic variables are also fixed by exact
operator identities:
\begin{align}
  T_1 = \langle\hat n\hat a\hat\sigma^z\rangle=0,\qquad
  T_2 = \langle\hat a\hat n\hat\sigma^z\rangle=U,\qquad
  C_n = \langle\hat n\hat a\hat\sigma^+\rangle=0,\qquad
  S_d = \langle\hat n\hat a^\dagger\hat\sigma^-\rangle=D.
\end{align}
In the numerical implementation $C_n=0$ and $S_d=D$ are imposed at each
integration step; $T_1$ and $T_2$ are propagated as free variables whose
dynamics automatically satisfy the identity-constrained values within
numerical tolerance.  To check that these restrictions are not merely cosmetic,
Fig.~\ref{fig:SI_CE4_identity_constraints} compares CE4 dynamics with and
without enforcing the strict one-excitation identities.  When the identities
are enforced, the photon number remains bounded and follows the expected
Rabi exchange.  If the quartic moments are allowed to evolve without these
operator constraints, the CE4 closure generates components outside the
two-state manifold $\{|1,g\rangle,|0,e\rangle\}$; the resulting photon number
rapidly becomes unphysical and the ODE integration fails.  This confirms that
the higher-order single-emitter CE hierarchy used here should be understood as
an identity-constrained benchmark for $N=1$, $N_\mathrm{exc}=1$, rather than as
a generic unconstrained high-order closure.\\

{\bf CE5.} CE5 extends the same hierarchy by promoting the quintic moments
\begin{align}
  L_2=\langle\hat a^2\hat n\hat\sigma^+\rangle,\qquad
  M_1=\langle\hat n^2\hat\sigma^-\rangle,
\end{align}
and closing the next generated order.  The numerical CE5 implementation also
evolves $x=\operatorname{artanh}(w)$ instead of $w$, so that the recovered
inversion $w=\tanh x$ remains bounded by construction.  CE3--CE5 are therefore
valid here only as implementations of the $N=1$, $N_\mathrm{exc}=1$ benchmark;
the single-excitation reductions above are not valid for generic coherent
states or for many-molecule TC dynamics.

\subsection{Many-molecule CE2 hierarchy}
\label{subsec:SI_many_molecule_CE2}

For $N>1$, a controlled many-molecule
CE2 hierarchy must retain intermolecular two-body moments; a single-emitter CE
hierarchy plus molecular MF factorization misses finite-$N$ pair correlations.
Under permutation symmetry, CE2 retains
\begin{align}
  \alpha=\langle\hat a\rangle,\quad
  m=\langle\hat a^2\rangle,\quad
  n=\langle\hat a^\dagger\hat a\rangle,\quad
  s=\langle\hat\sigma_i^-\rangle,\quad
  w=\langle\hat\sigma_i^z\rangle,
\end{align}
field--molecule moments
\begin{align}
  V=\langle\hat a\hat\sigma_i^-\rangle,\quad
  C=\langle\hat a\hat\sigma_i^+\rangle,\quad
  D=\langle\hat a^\dagger\hat\sigma_i^-\rangle,\quad
  U=\langle\hat a\hat\sigma_i^z\rangle,
\end{align}
and, for $i\ne j$, intermolecular moments
\begin{align}
  M=\langle\hat\sigma_i^-\hat\sigma_j^-\rangle,\quad
  P=\langle\hat\sigma_i^+\hat\sigma_j^-\rangle,\quad
  R=\langle\hat\sigma_i^-\hat\sigma_j^z\rangle,\quad
  Z=\langle\hat\sigma_i^z\hat\sigma_j^z\rangle .
\end{align}
Thus
\begin{align}
  \mathbf y_{\mathrm{CE2}}^{(N)}
  =
  (\alpha,m,n,s,w,V,C,D,U,M,P,R,Z).
\end{align}

The one-body and photon-autocorrelation equations are
\begin{align}
  \dot\alpha&=-ig_\mathrm{c}Ns,\\
  \dot s&=ig_\mathrm{c}U,\\
  \dot w&=2ig_\mathrm{c}(D-C),\\
  \dot n&=ig_\mathrm{c}N(C-D),\\
  \dot m&=-2ig_\mathrm{c}NV,
\end{align}
and conserve
\begin{align}
  N_\mathrm{exc}=n+\frac{N}{2}(1+w).
\end{align}
For compactness define the CE2-closed third moments
\begin{align}
  A_z
  &=
  nw+\alpha U^*+\alpha^*U-2|\alpha|^2w,\\
  A_z^+
  &=
  (n+1)w+\alpha U^*+\alpha^*U-2|\alpha|^2w,\\
  B_-
  &=
  (n+1)s+\alpha^*V+\alpha D-2|\alpha|^2s,\\
  B_+
  &=
  ms^*+2\alpha C-2\alpha^2s^*,\\
  A_{2z}
  &=
  mw+2\alpha U-2\alpha^2w.
\end{align}
The field--molecule two-body equations are then
\begin{align}
  \dot V
  &=
  -ig_\mathrm{c}(N-1)M+ig_\mathrm{c}A_{2z},\\
  \dot C
  &=
  -ig_\mathrm{c}
  \left[
    A_z^+ + \frac{1-w}{2} + (N-1)P
  \right],\\
  \dot D
  &=
  ig_\mathrm{c}
  \left[
    A_z + \frac{1+w}{2} + (N-1)P^*
  \right],\\
  \dot U
  &=
  -ig_\mathrm{c}\left[s+(N-1)R\right]
  +2ig_\mathrm{c}B_-
  -2ig_\mathrm{c}B_+ .
\end{align}
The explicit $(N-1)M$, $(N-1)P$, and $(N-1)R$ terms are the finite-$N$
intermolecular corrections absent from the single-emitter hierarchy.

The intermolecular equations require one-field--two-spin third moments.  With
CE2 closure,
\begin{align}
  X_R&=\alpha R+Us+Vw-2\alpha ws,\\
  X_+&=\alpha R^*+Cw+Us^*-2\alpha ws^*,\\
  Y_-&=\alpha^*R+U^*s+Dw-2\alpha^*ws,\\
  X_Z&=\alpha Z+2wU-2\alpha w^2,\\
  Y_M&=\alpha^*M+2Ds-2\alpha^*s^2,\\
  X_P&=\alpha P^*+Vs^*+Cs-2\alpha|s|^2 .
\end{align}
The closed CE2 equations are
\begin{align}
  \dot M&=2ig_\mathrm{c}X_R,\\
  \dot P&=ig_\mathrm{c}(X_+-Y_-),\\
  \dot R&=ig_\mathrm{c}X_Z+2ig_\mathrm{c}Y_M-2ig_\mathrm{c}X_P,\\
  \dot Z&=4ig_\mathrm{c}(Y_--X_+).
\end{align}
For the many-molecule one-photon benchmark used to test this hierarchy, the
initial state is $|1\rangle\otimes|g_1\cdots g_N\rangle$, giving
\begin{align}
  &\alpha(0)=m(0)=s(0)=V(0)=C(0)=D(0)=U(0)=M(0)=P(0)=R(0)=0,\qquad \notag\\
  &n(0)=1,\quad w(0)=-1,\quad Z(0)=1 .
\end{align}

\subsection{Numerical details}
\label{subsec:CE_numerics}

All CE calculations in this section are run in the resonant rotating frame with
$\omega_\mathrm{c}=\omega_0=2.0$~eV, over $t\in[0,200]$~fs.  The ODEs are
integrated with \texttt{scipy.integrate.solve\_ivp} (\texttt{RK45}) on $4001$
output points using \texttt{rtol}$=10^{-8}$, \texttt{atol}$=10^{-10}$ for
CE2--CE4, and tighter tolerances \texttt{rtol}$=10^{-9}$,
\texttt{atol}$=10^{-11}$ for CE5 and the many-molecule CE2 runs.
For the $N=1$, $N_\mathrm{exc}=1$ benchmark, the reference solution is the
two-state JC dynamics in $\{|1,g\rangle,|0,e\rangle\}$.  The MF curve uses a
coherent cavity initial condition with $|\alpha_0|^2=1$ while the CE
hierarchy is initialized from the Fock state $|1\rangle\otimes|g\rangle$; the
MF photon number is $|\alpha(t)|^2$.  For the
many-molecule CE2 tests, $g_\mathrm{c}$ denotes the single-molecule coupling;
when comparing different $N$ at fixed collective Rabi splitting we use
$g_\mathrm{c}=0.1/\sqrt{N}$~eV.

\newpage
\section{Bright-sector vibronic coupling, polaron decoupling, and SE/MF comparison}
\label{sec:SI_HTC_polaron}

This section supports the main-text HTC model simulations: in the overlap regime, SE and
MF give the same bright-sector linear dynamics, although they reach that limit
through different representations.

\subsection{Bright-dark basis transformation and bright-sector coupling}
\label{subsec:SI_polaron}

We derive the bright-sector projection of the HTC Hamiltonian using a discrete
Fourier basis for the molecular operators.

\paragraph*{Fourier mode definitions.}
Define the Fourier-transformed electronic and vibrational operators for $k=0,1,\ldots,N-1$:
\begin{equation}
  \hat{D}_k = \frac{1}{\sqrt{N}}\sum_{n=1}^{N}
              e^{\,2\pi i nk/N}\,\hat{\sigma}_n^{-},
  \qquad
  \hat{b}_k  = \frac{1}{\sqrt{N}}\sum_{n=1}^{N}
              e^{\,2\pi i nk/N}\,\hat{b}_n,
  \label{eq:SI_Fourier_def}
\end{equation}
with $[\hat{b}_k,\hat{b}_{k'}^\dagger]=\delta_{kk'}$.
In the low-excitation limit $\hat{D}_k$ satisfies $[\hat{D}_k,\hat{D}_{k'}^\dagger]\approx\delta_{kk'}$.
The $k=0$ modes are the collective bright operators,
\begin{equation}
  \hat{D}_0 \equiv \hat{B} = \frac{1}{\sqrt{N}}\sum_n\hat{\sigma}_n^-,
  \qquad
  \hat{b}_0 \equiv \hat{b}_B = \frac{1}{\sqrt{N}}\sum_n\hat{b}_n,
\end{equation}
and $k=1,\ldots,N-1$ label the $N-1$ dark modes.
The inverse transforms are
\begin{equation}
  \hat{\sigma}_n^- = \frac{1}{\sqrt{N}}\sum_k e^{-2\pi i nk/N}\hat{D}_k,
  \qquad
  \hat{b}_n = \frac{1}{\sqrt{N}}\sum_k e^{-2\pi i nk/N}\hat{b}_k.
  \label{eq:SI_Fourier_inv}
\end{equation}
The key identity used throughout is
\begin{equation}
  \frac{1}{N}\sum_{n=1}^{N} e^{\,2\pi i nq/N} = \delta_{q,0\!\!\pmod{N}}.
  \label{eq:SI_Fourier_ortho}
\end{equation}

\paragraph*{Light-matter coupling.}
Substituting Eq.~\ref{eq:SI_Fourier_inv} into the TC coupling:
\begin{align}
  g_\mathrm{c}\sum_n\!\left(\hat{a}^\dagger\hat{\sigma}_n^- + \mathrm{h.c.}\right)
  &= \frac{g_\mathrm{c}}{\sqrt{N}}\hat{a}^\dagger
     \sum_k\!\underbrace{\left(\sum_n e^{-2\pi i nk/N}\right)}_{=\,N\delta_{k,0}}
     \!\hat{D}_k + \mathrm{h.c.}
  = g_\mathrm{c}\sqrt{N}\!\left(\hat{a}^\dagger\hat{B}+\hat{a}\hat{B}^\dagger\right).
  \label{eq:SI_LM_Fourier}
\end{align}
All dark modes ($k\neq 0$) decouple from the cavity exactly by Eq.~\ref{eq:SI_Fourier_ortho}.

\paragraph*{Holstein coupling and bright-sector projection.}
Insert Eq.~\ref{eq:SI_Fourier_inv} into
$\hat{H}_\mathrm{e\text{-}ph}=\hbar c_\nu\sum_n\hat{\sigma}_n^+\hat{\sigma}_n^-(\hat{b}_n^\dagger+\hat{b}_n)$.
Using
\begin{equation}
  \hat{\sigma}_n^+\hat{\sigma}_n^-
  = \frac{1}{N}\sum_{k,k'} e^{2\pi i n(k-k')/N}\hat{D}_k^\dagger\hat{D}_{k'},
  \qquad
  \hat{b}_n^\dagger+\hat{b}_n
  = \frac{1}{\sqrt{N}}\sum_q\!\left(e^{\,2\pi i nq/N}\hat{b}_q^\dagger
    +e^{-2\pi i nq/N}\hat{b}_q\right),
\end{equation}
and summing over $n$ with Eq.~\ref{eq:SI_Fourier_ortho}:
\begin{align}
  \hat{H}_\mathrm{e\text{-}ph}
  &= \frac{\hbar c_\nu}{\sqrt{N}}\sum_{k,k'=0}^{N-1}
     \hat{D}_k^\dagger\hat{D}_{k'}\!
     \left(\hat{b}_{k'-k}^\dagger + \hat{b}_{k-k'}\right),
  \label{eq:SI_Heph_Fourier}
\end{align}
where all phonon indices are taken mod $N$.
Isolating the $k=k'=0$ term gives the bright-sector Holstein coupling used in the
main text,
\begin{equation}
  \hat{H}_\mathrm{e\text{-}ph}\big|_{k=k'=0}
  = \frac{\hbar c_\nu}{\sqrt{N}}\hat{B}^\dagger\hat{B}(\hat{b}_B^\dagger+\hat{b}_B),
  \label{eq:SI_Heph_bright}
\end{equation}
with the $1/\sqrt{N}$ suppression relative to the bare coupling $c_\nu$.
All remaining terms ($k\neq 0$ or $k'\neq 0$) involve at least one dark mode and
constitute the dark-mode couplings noted in the main text.

\paragraph*{Full HTC Hamiltonian in the Fourier basis.}
Collecting all terms (and $\sum_n\hat{b}_n^\dagger\hat{b}_n=\sum_k\hat{b}_k^\dagger\hat{b}_k$):
\begin{align}
  \hat{H}_\mathrm{HTC}
  &= \hbar\omega_\mathrm{c}\hat{a}^\dagger\hat{a}
   + \hbar\omega_0\sum_{k=0}^{N-1}\hat{D}_k^\dagger\hat{D}_k
   + \hbar\nu\sum_{k=0}^{N-1}\hat{b}_k^\dagger\hat{b}_k
  + \hbar g_\mathrm{c}\sqrt{N}\!\left(\hat{a}^\dagger\hat{B}+\hat{a}\hat{B}^\dagger\right)
  + \frac{\hbar c_\nu}{\sqrt{N}}\sum_{k,k'}\hat{D}_k^\dagger\hat{D}_{k'}
    \!\left(\hat{b}_{k'-k}^\dagger+\hat{b}_{k-k'}\right).
  \label{eq:SI_HTC_Fourier}
\end{align}
The cavity field couples only to the bright mode ($k=0$); the Holstein interaction
mixes electronic mode $k$ with mode $k'$ by absorbing or emitting a phonon of
quasi-momentum $k'-k$.
Restricting to $k=k'=0$ (all dark modes unoccupied) recovers the
bright-sector Hamiltonian with vibronic coupling $c_\nu/\sqrt{N}$.

\paragraph*{Polaron decoupling in the bright manifold.}
Within the bright subspace ($\hat{D}_{k\neq 0}|\psi\rangle=0$), only the $k=k'=0$
Holstein term survives, and the effective vibronic coupling is
$c_\nu/\sqrt{N}$~\cite{Herrera_Spano_PRL,Ying_ARPC2026}, renormalizing the
reorganization energy from $\lambda=c_\nu^2/(2\nu)$ to $\lambda_N=\lambda/N$.
In the polariton eigenbasis at resonance ($\Theta=\pi/4$), the vibronic matrix element
coupling the zero-phonon upper polariton to the one-phonon lower polariton is
\begin{align}
  \langle\mathrm{LP},1_B|\hat{H}_\mathrm{e\text{-}ph}|\mathrm{UP},0_B\rangle
  = \frac{c_\nu}{\sqrt{N}}\cos\Theta\sin\Theta
  = \frac{c_\nu}{2\sqrt{N}},
  \label{eq:SI_polaron_matrix}
\end{align}
and the Fermi golden rule rate scales as $\Gamma_{\mathrm{UP}\to\mathrm{LP}+1_B}\propto 1/N$.
The polariton reorganization energy is $\lambda_N=\lambda/(4N)$ at
resonance~\cite{Herrera_Spano_PRL,Ying_ARPC2026}.

\subsection{Two-internal-state HTC model and the \texorpdfstring{$4\times 4$}{4x4} interaction matrix}
\label{subsec:SI_4x4}

We use the minimal two-internal-state model of Ref.~\citenum{Cui_JCP2022}.
The basis states are:
$|B_0\rangle = N^{-1/2}\sum_k|e_k,a_k\rangle\prod_{j\neq k}|g_j,a_j\rangle$,
$|B_1\rangle = N^{-1/2}\sum_k|e_k,b_k\rangle\prod_{j\neq k}|g_j,a_j\rangle$
(collective Frenkel excitations with $a$ and $b$ vibrational quantum, respectively), and
$|C_0\rangle = |1_\mathrm{ph}\rangle\otimes|G_0\rangle$,
$|C_1\rangle = |1_\mathrm{ph}\rangle\otimes N^{-1/2}\sum_k|g_k,b_k\rangle\prod_{j\neq k}|g_j,a_j\rangle$
(one-photon cavity states).
In this basis the Holstein vibronic coupling reads
\begin{equation}
\hat{H}_\mathrm{e\text{-}ph}
= \hbar c_\nu\sum_j |e_j\rangle\langle e_j|
  \bigl(|b_j\rangle\langle a_j| + |a_j\rangle\langle b_j|\bigr).
\label{eq:SI_Heph_explicit}
\end{equation}
Because the projector $|e_j\rangle\langle e_j|$ requires the molecule to be electronically excited, $\hat{H}_\mathrm{e\text{-}ph}$ annihilates any state in which all molecules are in $|g_j\rangle$.
In particular, $\hat{H}_\mathrm{e\text{-}ph}|C_0\rangle = \hat{H}_\mathrm{e\text{-}ph}|C_1\rangle = 0$.

In the ordered basis $\{|B_0\rangle,|B_1\rangle,|C_0\rangle,|C_1\rangle\}$, the full HTC Hamiltonian matrix is
\begin{align}
  [\hat{H}_\mathrm{HTC}]
  = \hbar
  \begin{pmatrix}
    \omega_0                     & c_\nu                  & g_\mathrm{c}\sqrt{N} & 0                   \\
    c_\nu                        & \omega_0 + \nu         & 0                    & g_\mathrm{c}        \\
    g_\mathrm{c}\sqrt{N}         & 0                      & \omega_\mathrm{c}    & 0                   \\
    0                            & g_\mathrm{c}           & 0                    & \omega_\mathrm{c} + \nu
  \end{pmatrix}.
  \label{eq:SI_HTC_4x4}
\end{align}
The light-matter couplings reveal the channel asymmetry: $\langle C_0|\hat{H}_\mathrm{LM}|B_0\rangle = \hbar g_\mathrm{c}\sqrt{N}$ ($a$-channel, collectively enhanced) and $\langle C_1|\hat{H}_\mathrm{LM}|B_1\rangle = \hbar g_\mathrm{c}$ ($b$-channel, not enhanced).
The Holstein coupling gives $\langle B_0|\hat{H}_\mathrm{e\text{-}ph}|B_1\rangle = \hbar c_\nu$, while $\langle C_0|\hat{H}_\mathrm{e\text{-}ph}|C_1\rangle = 0$ since neither $|C_0\rangle$ nor $|C_1\rangle$ carries an electronic excitation.

\subsection{Limiting cases and convergence of the two frameworks}
\label{subsec:SI_limiting_cases}

\subsubsection{SE large-\texorpdfstring{$N$}{N} limit: decoupling of \texorpdfstring{$|C_1\rangle$}{C1} and effective \texorpdfstring{$3\times 3$}{3x3} dynamics}
\label{subsubsec:SI_SE_largeN}

Inspecting the $4\times 4$ Hamiltonian in Eq.~\ref{eq:SI_HTC_4x4}, the two light--matter matrix elements are
\begin{equation}
  \langle C_0|\hat{H}_\mathrm{LM}|B_0\rangle = \hbar g_\mathrm{c}\sqrt{N},
  \qquad
  \langle C_1|\hat{H}_\mathrm{LM}|B_1\rangle = \hbar g_\mathrm{c}.
  \label{eq:SI_LM_asymmetry}
\end{equation}
The natural energy unit of the problem is set by the collectively enhanced coupling $\hbar g_\mathrm{c}\sqrt{N}$.
On this scale the $b$-channel matrix element $\hbar g_\mathrm{c}=\hbar g_\mathrm{c}\sqrt{N}\cdot N^{-1/2}$ is suppressed by $1/\sqrt{N}$ and vanishes as $N\to\infty$.
Because $|C_1\rangle$ is connected to the rest of the basis only through this bare coupling (the vibronic Hamiltonian annihilates $|C_1\rangle$ identically, see Eq.~\ref{eq:SI_Heph_explicit}), the state $|C_1\rangle$ gradually decouples as $N$ grows.
In the strict $N\to\infty$ limit the dynamics is therefore confined to the subspace $\{|B_0\rangle,|B_1\rangle,|C_0\rangle\}$, governed by the effective $3\times 3$ Hamiltonian
\begin{equation}
  H^{(3)}_\mathrm{SE}
  = \hbar
  \begin{pmatrix}
    \omega_0               & c_\nu              & g_\mathrm{c}\sqrt{N} \\
    c_\nu                  & \omega_0+\nu       & 0                    \\
    g_\mathrm{c}\sqrt{N}   & 0                  & \omega_\mathrm{c}
  \end{pmatrix},
  \quad
  \text{basis: }\{|B_0\rangle,|B_1\rangle,|C_0\rangle\}.
  \label{eq:SI_H3_SE}
\end{equation}
The off-diagonal $c_\nu$ element drives vibronic exchange between $|B_0\rangle$ and $|B_1\rangle$ at the bare Huang--Rhys rate, while the photon $|C_0\rangle$ couples to $|B_0\rangle$ with the collectively enhanced Rabi coupling $g_\mathrm{c}\sqrt{N}$.
As $N$ increases, the dominant dynamics is a slow vibronic exchange within the
$\{|B_0\rangle,|B_1\rangle\}$ pair, dressed by the large Rabi coupling to
$|C_0\rangle$.

\subsubsection{MF weak-excitation limit: effective three-state dynamics and coincidence with SE}
\label{subsubsec:SI_MF_weakexit}

We now show that the MF EOM in Eqs.~\ref{eq:SI_alpha_2fock} and~\ref{eq:SI_amp_2fock} reduce to the same $3\times 3$ Hamiltonian dynamics in the weak-excitation ($n_0\to 0$) limit, after an appropriate rescaling of $\alpha$.

Consider the single-photon-type initial condition
\begin{equation}
  \alpha(0)=1,\quad
  c_{ga}(0)=1,\quad
  c_{ea}(0)=c_{gb}(0)=c_{eb}(0)=0,\quad
  N\to\infty.
  \label{eq:SI_IC_dilute}
\end{equation}
In this limit $c_{ga}\approx 1$ dominates, while the excited-state amplitudes are small.
The amplitude $c_{gb}$ is one order smaller than $c_{ea}$ and $c_{eb}$ because it is generated only indirectly through the three-step chain
\begin{equation}
  |g,a\rangle \xrightarrow{g_\mathrm{c}\alpha} |e,a\rangle \xrightarrow{c_\nu} |e,b\rangle \xrightarrow{g_\mathrm{c}\alpha^*} |g,b\rangle,
  \label{eq:SI_cgb_chain}
\end{equation}
so $c_{gb}$ is second order in the small excitation amplitude.
Concretely, with the dilute scaling
\begin{equation}
  c_{ga}=1+O(N^{-1}),
  \qquad
  c_{ea},\,c_{eb}=O(N^{-1/2}),
  \qquad
  c_{gb}=O(N^{-1}),
  \label{eq:SI_dilute_scaling}
\end{equation}
the cross terms in Eq.~\ref{eq:SI_alpha_2fock} satisfy
\begin{equation}
  c_{ga}^*c_{ea}=c_{ea}+O(N^{-3/2}),\qquad
  c_{gb}^*c_{eb}=O(N^{-3/2}),
\end{equation}
so the cavity equation simplifies to
\begin{equation}
  \dot{\alpha} = -i\omega_\mathrm{c}\,\alpha - ig_\mathrm{c}N\,c_{ea} + O(N^{-1/2}).
  \label{eq:SI_alpha_3state}
\end{equation}
Similarly, setting $c_{ga}\to 1$ and $c_{gb}\to 0$ in Eqs.~\ref{eq:SI_cea} and~\ref{eq:SI_ceb} gives
\begin{align}
  \dot{c}_{ea} &= -i\omega_0\,c_{ea} - ic_\nu\,c_{eb} - ig_\mathrm{c}\,\alpha,
  \label{eq:SI_cea_3state}\\
  \dot{c}_{eb} &= -i(\omega_0+\nu)\,c_{eb} - ic_\nu\,c_{ea}.
  \label{eq:SI_ceb_3state}
\end{align}
Eqs.~\ref{eq:SI_alpha_3state}--\ref{eq:SI_ceb_3state} form a closed linear system in $(\alpha,c_{ea},c_{eb})$.
The coupling between $\alpha$ and $c_{ea}$ is $g_\mathrm{c}N$ vs.\ $g_\mathrm{c}$, which is asymmetric in the original variables.
Introducing the rescaled cavity amplitude
\begin{equation}
  \tilde{\alpha} \equiv \frac{\alpha}{\sqrt{N}},
  \label{eq:SI_alpha_rescale}
\end{equation}
all three equations become symmetric in the coupling:
\begin{align}
  \dot{\tilde{\alpha}} &= -i\omega_\mathrm{c}\,\tilde{\alpha} - ig_\mathrm{c}\sqrt{N}\,c_{ea},
  \label{eq:SI_alpha_tilde}\\
  \dot{c}_{ea} &= -ig_\mathrm{c}\sqrt{N}\,\tilde{\alpha} - i\omega_0\,c_{ea} - ic_\nu\,c_{eb},
  \label{eq:SI_cea_tilde}\\
  \dot{c}_{eb} &= -ic_\nu\,c_{ea} - i(\omega_0+\nu)\,c_{eb}.
  \label{eq:SI_ceb_tilde}
\end{align}
These are precisely the Schr\"{o}dinger equations
$i\partial_t\mathbf{v}=H^{(3)}_\mathrm{SE}\,\mathbf{v}$ generated by
Eq.~\ref{eq:SI_H3_SE}, with the state vector
\begin{equation}
  \mathbf{v}(t) = \bigl(\tilde{\alpha}(t),\,c_{ea}(t),\,c_{eb}(t)\bigr)^T
  \;\longleftrightarrow\;
  \bigl(|C_0\rangle,\,|B_0\rangle,\,|B_1\rangle\bigr).
  \label{eq:SI_3state_identification}
\end{equation}
Hence, after the rescaling $\alpha\to\tilde{\alpha}=\alpha/\sqrt{N}$, the MF
equations in the $n_0\to 0$ limit reduce to the same $3\times 3$ linear system as
the SE equations in the large-$N$ limit. The two frameworks therefore share the same
bright-sector dynamics in the overlap regime; differences arise from finite-$N$
corrections on the SE side and finite-$n_0$ nonlinear corrections on the MF side.

\subsection{Relation between TDH and SE approximations}
\label{subsec:SI_TDH_SE}

It is useful to distinguish agreement of observables in the overlap regime from
equality of wavefunction structure. For $N$ two-level molecules with amplitudes
$c_{jg}$ and $c_{je}$, the TDH product state expands as
\begin{align}
  \prod_{j} |\psi_j(t)\rangle
  &= \prod_j\!\bigl(c_{jg}|g_j\rangle + c_{je}|e_j\rangle\bigr)
  \notag\\
  &= \prod_j\!c_{jg}\,|G\rangle
   + \sum_k c_{ke}\!\!\prod_{j\neq k}\!\!c_{jg}\,
     |e_k\rangle\!\!\prod_{j\neq k}\!\!|g_j\rangle
   + \cdots.
  \label{eq:SI_TDH_expand}
\end{align}
Under permutation symmetry ($c_{je}=c_e$, $c_{jg}\approx c_g$), the single-excitation sector reduces to
\begin{equation}
  c_e\,c_g^{N-1}\sqrt{N}\,|B\rangle,
\end{equation}
confirming that the single-excitation weight resides entirely on the bright state $|B\rangle$.
This is why the permutation-symmetric MF equations involve only
$\alpha \equiv \langle\hat{a}\rangle$ and
$\sigma \equiv N^{-1}\sum_n\langle\hat{\sigma}_n^-\rangle$: these variables track the
cavity amplitude and the bright-exciton amplitude.
TDH/MF and SE can therefore share the same bright-sector linear equations in the
overlap regime. However, TDH does not represent the same entangled light--matter wavefunction
structure as SE.

\bibliography{ref}